\pdfoutput=1

\documentclass[twocolumn, 11pt]{aastex7}

\makeatletter
\let\frontmatter@title@above=\relax
\makeatother

\usepackage{natbib}
\let\tablenum\relax
\usepackage{siunitx}
\usepackage{tabularx}
\usepackage{booktabs}
\usepackage{multirow}
\usepackage{lmodern}
\usepackage{slantsc}
\usepackage{scrextend}
\usepackage{caption}
\usepackage{subcaption}
\usepackage{etoolbox}
\usepackage{xfrac}
\pretocmd{\abstractname}{\newpage}{}{}
\usepackage[style=american]{csquotes}
\newcommand{\toinum}{\mbox{\textsc{toi}--1232}}
\newcommand{\toinumb}{\toinum\,b}
\newcommand{\toinumc}{\toinum\,c}
\newcommand{\toinumbc}{\toinum\,b\,\&\,c}
\newcommand{\ticnum}{\textsc{tic}\,364395227}
\newcommand{\gaianum}{\emph{Gaia} \textsc{dr}2~5289147741358156928}
\newcommand{\gaia}{\textit{Gaia}}
\newcommand{\ttv}{\textsc{ttv}} % defined
\newcommand{\feros}{\textsc{feros}} % defined
\newcommand{\tess}{\textsl{\textsc{tess}}} % defined
\newcommand{\wine}{\textsc{wine}} % defined
\newcommand{\rv}{\textsc{rv}} % defined
\newcommand{\mmr}{\textsc{mmr}} % defined
\newcommand{\bjd}{\textsc{bjd}}
\newcommand{\fwhm}{\textsc{fwhm}} % defined
\newcommand{\bis}{\textsc{bis}}
\newcommand{\ld}{\textsc{ld}}

\newcommand{\symba}{\textsc{s}y\textsc{mba}}
\newcommand{\gastli}{\textsc{gastli}} % defined
\newcommand{\mesa}{\textsc{mesa}}
\newcommand{\tpf}{\textsc{tpf}}
\newcommand{\nasa}{\textsc{nasa}}
\newcommand{\spoc}{\textsc{spoc}}
\newcommand{\sap}{\textsc{sap}}
\newcommand{\parsecabbr}{\textsc{parsec}}
\newcommand{\ffi}{\textsc{ffi}}
\newcommand{\pdc}{\textsc{pdc}}
\newcommand{\ccd}{\textsc{ccd}}
\newcommand{\utc}{\textsc{utc}}
\newcommand{\tls}{\textsc{tls}}
\newcommand{\gls}{\textsc{gls}}
\newcommand{\ns}{\textsc{ns}}
\newcommand{\mcmc}{\textsc{mcmc}}

\newcommand{\lcogt}{\textsc{lcogt}}
\newcommand{\sso}{\textsc{sso}}
\newcommand{\soar}{\textsc{soar}}

\newcommand{\astep}{\textsc{astep}}
\newcommand{\bnsf}{\textsc{bnsf}}
\newcommand{\vihren}{\textsc{vihren}}

\newcommand{\tesseract}{\texttt{tesseract}}
\newcommand{\lightkurve}{\texttt{lightkurve}}

\newcommand{\derived}{(\textsc{derived})}
\newcommand{\thispap}{\textsc{this paper}}

\newcommand{\Mjup}{M_{\rm{Jup}}}
\newcommand{\incl}{i}

\newcolumntype{Y}{>{\raggedright\arraybackslash}X}

\DeclareSIUnit\year{yr}
\DeclareSIUnit\mag{mag}
\DeclareSIUnit\parsec{pc}
\DeclareSIUnit\arcs{arcsec}
\DeclareSIUnit{\days}{days}
\DeclareSIUnit{\pixel}{pixel}
\DeclareSIUnit{\kcount}{K}
\DeclareSIUnit[unit-font-command=\mathit]{\Msun}{M_{\odot}}
\DeclareSIUnit[unit-font-command=\mathit]{\Rsun}{R_{\odot}}
\DeclareSIUnit[unit-font-command=\mathit]{\MJup}{M_{\rm{Jup}}}
\DeclareSIUnit[unit-font-command=\mathit]{\RJup}{R_{\rm{Jup}}}
\DeclareSIUnit[unit-font-command=\mathit]{\MEarth}{M_{\oplus}}
\DeclareSIUnit[unit-font-command=\mathit]{\REarth}{R_{\oplus}}

\let\orgautoref\autoref
\renewcommand{\autoref}{%
    \def\equationautorefname{Eq.}%
    \def\figureautorefname{Fig.}%
    \def\sectionautorefname{Sec.}%
    \def\subsectionautorefname{Sec.}%
    \def\subsubsectionautorefname{Sec.}%
    \orgautoref
}

\begin{document}

\title{A warm massive pair of planets around TOI--1232 revealed with transit timing variations and Doppler spectroscopy\footnote{Based on observations collected at the European Organization for Astronomical Research in the Southern Hemisphere under MPG programmes 0102.A-9006, 0103.A-9008, 0104.A-9007.}}

\newcommand\cerca{CERCA/ISO, Department of Physics, Case Western Reserve University, 10900 Euclid Avenue, Cleveland, Ohio 44106, USA}
\newcommand\mpaheid{Max-Planck-Institut f\"{u}r Astronomie, K\"{o}nigstuhl 17, D-69117 Heidelberg, Germany}
\newcommand\landwart{Landessternwarte, Zentrum f\"ur Astronomie der Universt\"at Heidelberg, K\"onigstuhl 12, 69117 Heidelberg, Germany}
\newcommand\susofia{Department of Astronomy, Faculty of Physics, Sofia University ``St Kliment Ohridski'', 5 James Bourchier Blvd, 1164 Sofia, Bulgaria}
\newcommand\ohiost{Department of Astronomy, McPherson Laboratory, The Ohio State University, 140 W 18th Ave, Columbus, Ohio 43210, USA}
\newcommand\facultychile{Facultad de Ingeniera y Ciencias, Universidad Adolfo Ib\'{a}\~{n}ez, Av. Diagonal las Torres 2640, Pe\~{n}alol\'{e}n, Santiago, Chile}
\newcommand\instchile{Millennium Institute for Astrophysics, Camino El Observatorio 1515, Las Condes, Santiago, Chile}
\newcommand\pontchile{Instituto de Astrof\'isica, Facultad de F\'isica, Pontificia Universidad Cat\'olica de Chile, Chile}
\newcommand\arizona{Department of Astronomy/Steward Observatory, The University of Arizona, 933 North Cherry Avenue, Tucson, AZ 85721, USA}
\newcommand\stsi{Space Telescope Science Institute, Steven Muller Building, 3700 San Martin Drive, Baltimore, MD 21218, USA}
\newcommand\lund{Lund Observatory, Division of Astrophysics, Department of Physics, Lund University, Box 43, 22100 Lund, Sweden}
\newcommand\rasuk{Royal Astronomical Society, Burlington House, Piccadilly, London W1J 0BQ, UK}
\newcommand\obscotedazur{Observatoire de la C\^ote d'Azur, Laboratoire Lagrange, CNRS, Boulevard de l'Observatoire, CS 34229, 06304 Nice cedex 04, France}
\newcommand\bham{School of Physics \& Astronomy, University of Birmingham, Edgbaston, Birmingham B15 2TT, United Kingdom}
\newcommand\pnraipev{Programma Nazionale di Ricerche in Antartide, Institut polaire français Paul-Émile Victor, Concordia Station, Antarctica}
\newcommand\ens{\'Ecole Normale Sup\'erieure, D\'epartement de Physique, Rue d'Ulm, 75005 Paris cedex 5, France}
\newcommand\eso{European Southern Observatory, Alonso de C\'ordova 3107, Vitacura, Casilla 19001, Santiago, Chile}
\newcommand\esogermany{European Southern Observatory, Karl-Schwarzschild-Straße 2, 85748 Garching bei München, Germany}
\newcommand\texas{Department of Physics, Engineering and Astronomy, Stephen F. Austin State University, 1936 North St, Nacogdoches, TX 75962, USA}
\newcommand\kavlimit{Department of Physics and Kavli Institute, MIT, 77 Massachusetts Avenue, Cambridge, MA 02139, USA}
\newcommand\cfa{Harvard -- Smithsonian Center for Astrophysics, 60 Garden Street, Cambridge, MA 02138, USA}
\newcommand\princeton{Department of Astrophysical Sciences, Peyton Hall, 4 Ivy Lane, Princeton University, Princeton, NJ 08544, USA}
\newcommand\deapsmit{Department of Earth, Atmospheric and Planetary Sciences, MIT, 77 Massachusetts Avenue, Cambridge, MA 02139, USA}
\newcommand\daamit{Department of Aeronautics and Astronautics, MIT, 77 Massachusetts Avenue, Cambridge, MA 02139, USA}
\newcommand\nasames{NASA Ames Research Center, Moffett Field, CA 94035, USA}
\newcommand\nasagod{NASA Goddard Space Flight Center, 8800 Greenbelt Rd, Greenbelt, MD 20771, USA}
\newcommand\aavso{American Association of Variable Star Observers, 49 Bay State Road, Cambridge, MA 02138, USA}
\newcommand\hazelwood{Hazelwood Observatory, Churchill, Victoria 3842, Australia}
\newcommand\tsinghua{Department of Astronomy, Westlake University, Hangzhou 310030, Zhejiang Province, China}
\newcommand\cavendish{Cavendish Laboratory, Department of Physics, University of Cambridge, J J Thomson Avenue, Cambridge, CB3 0HE, UK}
\newcommand\elsauce{Obstech -- El Sauce Observatory, Río Hurtado, Coquimbo Province, Chile}
\newcommand\geneve{Observatoire Astronomique de l’Université de Genève, Chemin Pegasi 51b, 1290 Versoix, Switzerland}
\newcommand\tenerifedept{Departamento de Astrof\'isica, Universidad de La Laguna (ULL), Av. Astrofisico Francisco Sánchez, E-38206 La Laguna, Tenerife, Spain}
\newcommand\tenerifeobs{Instituto de Astrof\'isica de Canarias (IAC), C. Vía Láctea, E-38200 La Laguna, Tenerife, Spain}
\newcommand\seti{SETI Institute, 339 North Bernardo Ave, Suite 200, Mountain View, CA 94043, USA}

\correspondingauthor{\\Deyan P. Mihaylov,\\\href{mailto:me@deyanmihaylov.net}{me@deyanmihaylov.net}}

% Deyan P. Mihaylov <deyan.mihaylov@case.edu>
\author[0000-0002-8820-407X]{Deyan P. Mihaylov}
\affil{\cerca}
\affil{\susofia}
\email{deyan.mihaylov@case.edu}

% Jan Eberhardt <eberhardt@mpia.de>
\author[0000-0003-3130-2768]{Jan Eberhardt}
\affil{\mpaheid}
\email{eberhardt@mpia.de}

% Trifon Trifonov <trifonov@mpia.de>
\author[0000-0002-0236-775X]{Trifon Trifonov}
\affil{\mpaheid}
\affil{\landwart}
\affil{\susofia}
\email{trifonov@mpia.de}

% Rafael Brahm <rbrahm@gmail.com>
\author[0000-0002-9158-7315]{Rafael Brahm}
\affil{\facultychile}
\affil{\instchile}
\email{rafael.brahm@uai.cl}

% Thomas Henning <henning@mpia.de>
\author[0000-0002-1493-300X]{Thomas Henning}
\affil{\mpaheid}
\email{henning@mpia.de}

% Andres Jordan <andres.jordan@gmail.com>
\author[0000-0002-5389-3944]{Andr\'{e}s Jord\'{a}n}
\affil{\facultychile}
\affil{\instchile}
\affil{\elsauce}
\email{andres.jordan@uai.cl}

% Denitza Stoeva <dstoeva@phys.uni-sofia.bg>
\author[0000-0001-6277-9644]{Denitza Stoeva}
\affil{\susofia}
\email{dstoeva@phys.uni-sofia.bg}

% Matias Jones <mjones@eso.org>
\author{Mat\'ias I. Jones}
\affiliation{\eso}
\email{mjones@eso.org}

% Lorena Acuna-Aguirre <acuna@mpia.de>
\author[0000-0002-9147-7925]{Lorena Acu\~{n}a-Aguirre}
\affiliation{\mpaheid}
\email{acuna@mpia.de}

% Stefan Stefanov <sstefanov@nao-rozhen.org>
\author[0000-0002-4993-2840]{Stefan Stefanov}
\affil{\susofia}
\email{sstefanov@nao-rozhen.org}

% Marcelo Tala Pinto <marcelo.tala@edu.uai.cl>
\author[0009-0004-8891-4057]{M. Tala Pinto} 
\affil{\ohiost}
\email{marcelo.tala@edu.uai.cl}

% Melissa Hobson <melissa.hobson@unige.ch>
\author[0000-0002-5945-7975]{Melissa J. Hobson}
\affil{\geneve}
\email{melissa.hobson@unige.ch}

% Nestor Espinoza <nespinoza@stsci.edu>
\author[0000-0001-9513-1449]{Nestor Espinoza}
\affil{\stsi}
\email{nespinoza@stsci.edu}

% Felipe Rojas <firojas@uc.cl>
\author[0000-0003-3047-6272]{Felipe I. Rojas}
\affil{\pontchile}
\affil{\instchile}
\email{firojas@uc.cl}

% Martin Schlecker <martin.schlecker@eso.org>
\author[0000-0001-8355-2107]{Martin Schlecker}
\affil{\esogermany}
\affil{\arizona}
\email{martin.schlecker@eso.org}

% Vladimir Bozhilov <vbozhilov@phys.uni-sofia.bg>
\author[0000-0002-3117-7197]{Vladimir Bozhilov}
\affil{\susofia}
\email{vbozhilov@phys.uni-sofia.bg}

% Tristan Guillot <tristan.guillot@oca.eu>
\author[0000-0002-7188-8428]{Tristan Guillot}
\affil{\obscotedazur}
\email{tristan.guillot@oca.eu}

% Amaury Triaud <a.triaud@bham.ac.uk>
\author[0000-0002-5510-8751]{Amaury H. M. J. Triaud}
\affil{\bham}
\email{a.triaud@bham.ac.uk}

% Jack Lissauer <jack.lissauer@nasa.gov>
\author[0000-0001-6513-1659]{Jack J. Lissauer}
\affil{\nasames}
\email{jack.lissauer@nasa.gov}

% Judith Korth <judithkorth@googlemail.com>
\author[0000-0002-0076-6239]{Judith Korth}
\affil{\lund}
\affil{\geneve}
\email{judith.korth@fysik.lu.se}

% Hannu Parviainen <hpparvi@gmail.com>
\author[0000-0001-5519-1391]{Hannu Parviainen}
\altaffiliation{Ram\'on y Cajal Fellow}
\affil{\tenerifedept}
\affil{\tenerifeobs}
\email{hannu@iac.es}

% Laura Kreidberg <kreidberg@mpia.de>
\author[0000-0003-0514-1147]{Laura Kreidberg}
\affiliation{\mpaheid}
\email{kreidberg@mpia.de}

% Philippe Bendjoya <philippe.bendjoya@oca.eu>
\author[0000-0002-4278-1437]{Philippe Bendjoya}
\affil{\obscotedazur}
\email{philippe.bendjoya@oca.eu}

% Olga Suarez <olga.suarez@oca.eu>
\author[0000-0002-3503-3617]{Olga Suarez}
\affil{\obscotedazur}
\email{olga.suarez@oca.eu}

% Carl Ziegler <carl.ziegler@sfasu.edu>
\author[0000-0002-0619-7639]{Carl Ziegler}
\affiliation{\texas}
\email{carl.ziegler@sfasu.edu}

% Pamela Rowden <pam.rowden27@gmail.com>
\author[0000-0002-4829-7101]{Pamela Rowden}
\affil{\rasuk}
\email{pam.rowden27@gmail.com}

% Alexander Rudat <arudat@space.mit.edu>
\author{Alexander Rudat}
\affil{\kavlimit}
\email{arudat@space.mit.edu}

% Veselin Kostov <veselin.b.kostov@nasa.gov>
\author[0000-0001-9786-1031]{Veselin Kostov}
\affil{\nasagod}
\affil{\seti}
\email{veselin.b.kostov@nasa.gov}

% Joshua Winn <jnwinn@princeton.edu>
\author[0000-0002-4265-047X]{Joshua N. Winn}
\affil{\princeton}
\email{jnwinn@princeton.edu}

% Jon M. Jenkins <jon.jenkins@nasa.gov>
\author[0000-0002-4715-9460]{Jon M. Jenkins}
\affil{\nasames}
\email{jon.jenkins@nasa.gov}

% Karen Collins <karen.collins@cfa.harvard.edu> (LCO scheduling, data reduction, time contribution)
\author[0000-0001-6588-9574]{Karen A. Collins}
\affil{\cfa}
\email{karen.collins@cfa.harvard.edu}

% Cristilyn Watkins <cristilynwatkins@gmail.com> (SG1 coordination, LCO data reduction, write-up)
\author[0000-0001-8621-6731]{Cristilyn N. Watkins}
\affil{\cfa}
\email{cristilynwatkins@gmail.com}

% Don Radford <dradford63@gmail.com> (Brierfield observations and data reduction)
\author{Don J. Radford}
\affil{\aavso}
\email{dradford63@gmail.com}

% Chris Stockdale <thestockdalefamily@bigpond.com> (Hazelwood observations and data reduction)
\author[0000-0003-2163-1437]{Chris Stockdale}
\affil{\hazelwood}
\email{thestockdalefamily@bigpond.com}

% Tianjun Gan <tjgan@foxmail.com> (LCO data reduction)
\author[0000-0002-4503-9705]{Tianjun Gan}
\affil{\tsinghua}
\email{tjgan@foxmail.com}

\begin{abstract}
\noindent
TOI--1232 is a G--dwarf star with a mass of $1.06_{-0.06}^{+0.07} M_\odot$, a radius of $1.07\pm 0.05 R_\odot$, and  slightly higher metallicity than solar of Fe/H = $0.18 \pm 0.05$. 
The star hosts a transiting warm Jovian-mass planet, TOI--1232\,b, with an orbital period of $P_{b} = 14.256_{-0.001}^{+0.001}$ days, identified with data from multiple sectors of the \textit{TESS} space telescope.
The \textit{TESS} light curve of TOI--1232 is complex, as it is contaminated by a background eclipsing binary with a period of $1.37$ days. The TOI--1232\,b was firmly confirmed by ground-based transit follow-up campaigns from Las Cumbres, Hazelwood, Brierfield, and ASTEP observatories.
Additionally, the \textit{TESS} transits of TOI-1232\,b exhibit strong transit timing variations (TTVs) with a super-period of $235.5 \pm 0.7$ days and a semi-amplitude of 27 minutes. 
Radial velocity (RV) follow-up with the FEROS spectrograph confirms the planetary nature of the transiting candidate, while a self-consistent $N$-body analysis of RVs and TTVs pinpoints the presence of a second outer Saturn-mass companion, TOI--1232\,c with a period of $P_{c} = 30.356_{-0.012}^{+0.010}$ days. 
The TOI--1232 warm-giant system is particularly important due to the evidence of two massive planets that reside near the 2:1 commensurability but are not locked in a mean motion resonance (MMR). 
Thanks to \textit{TESS}, we have revealed a handful of these rare systems. 
Hence, TOI--1232 is an important addition to understanding the formation and dynamical evolution of such compact, massive, warm giant planets.
\end{abstract}


\section{Introduction}
\label{sec1}
\noindent
The past three decades of active exoplanet searches have confirmed more than 6,000 planets.\footnote{A continually updated list of confirmed exoplanets is available at \href{https://exoplanetarchive.ipac.caltech.edu/}{exoplanetarchive.ipac.caltech.edu}} 
Many of these exoplanets inhabit multiplanetary systems whose physical and orbital parameters have been investigated, providing critical insights into planet formation and long-term dynamical evolution. 
In contrast with our solar system, we have observed diverse physical and orbital characteristics in these systems, which suggest possible multiple pathways of planet formation and migration.

Warm Jovian-mass planets (with orbital periods between 10 and 300 days) can act as an important test bench for studying planetary migration mechanisms \citep[][]{Raymond2022}. 
Planets of this kind likely form in the colder, ice- and gas-rich regions of the protoplanetary disk, beyond the snow line, where rapid solid accretion is believed to be efficient during the first \(\sim 10\) million years of a system's evolution \citep[see, e.g.,][]{Dawson2018, Emsenhuber2021}. 
Understanding the migration of these planets inward from their formation regions provides critical tests of planet formation theories, particularly for systems where warm Jupiters are near or locked in low-order mean-motion resonances (\mmr). 
Such systems highlight the dynamic interactions between planets and the protoplanetary disk \citep[][]{Ida2010, Kley2012, Coleman2014, Bitsch2020, Schlecker2020b, Matsumura2021}.

Two primary migration mechanisms have been proposed to explain the presence of warm Jovian planets. 
The first mechanism is disk-driven migration, in which interactions between a planet and its protoplanetary disk induce an inward 
drift \citep[e.g.,][]{Lin1986}. 
The second mechanism, high-eccentricity tidal migration, involves dynamical instabilities that lead to planet-planet scattering and secular interactions, thus driving the planet to high orbital eccentricity, followed by circularisation by tides in warmer regions \citep[e.g.,][]{Rasio1996, Fabrycky2007, Bitsch2020}. 
Distinguishing between these mechanisms requires detailed statistical characterization of warm Jovian multiple-planet systems whose dynamics retain important signatures of their 
evolutionary history.

Recent studies have emphasized the role of resonances and orbital architectures in constraining the dominant migration pathways. 
For instance, the discovery of resonant chains in multiplanetary systems supports the hypothesis of smooth disk migration, while isolated warm Jupiters with eccentric orbits may favor high-eccentricity pathways \citep[][]{Huang2016, Petrovich2016, Santerne2016, Dong2021}. 
The continued discovery and characterization of warm Jovian systems will provide key observational constraints (e.g., eccentricity, alignments, and masses) on these competing theories and offer a deeper understanding of the diversity of planetary systems.

\begin{figure*}[tp]
\centering
\includegraphics[scale=1]{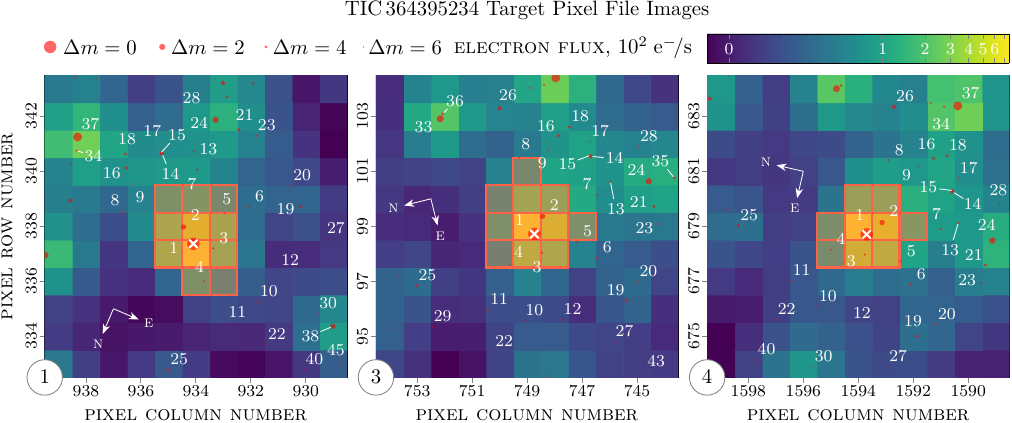}
    \caption{Target pixel file (\tpf) image of \toinum\ in \tess\ Sectors 1, 3, and 4. The remaining \tpf\ plots for \toinum\ are shown in Figs. \ref{fig:tpf2} and \ref{fig:tpf3}. The red dots represent the position of the \gaia\ sources in the field. \toinum\ resides in the middle, marked with a white \(\times\). The pixels marked with red borders are the ones used to construct the \tess\ Simple Aperture Photometry (\sap).}
\label{TPFnew} 
\end{figure*}

Accurately measuring the dynamical masses and orbital eccentricities of warm Jovian planets is one of the important tasks in modern exoplanet research. 
Specifically, \nasa's \textit{Transiting Exoplanet Survey Satellite} \citep[\tess,][]{Ricker2015} has enabled the detection of a large number of warm Jovian-mass planets around relatively bright stars, which are ideal for precise Doppler follow-up to determine planetary mass. 
When combined with radius measurements, these observations allow us to calculate the planets' bulk density, making them prime targets for transit spectroscopy to study exoplanet atmospheres using ground-based Doppler spectroscopy and the James Webb Space Telescope (\textsc{jwst}).

\begin{figure*}[t!]
    \centering
    \begin{subfigure}{\textwidth}
        \includegraphics[scale=1]{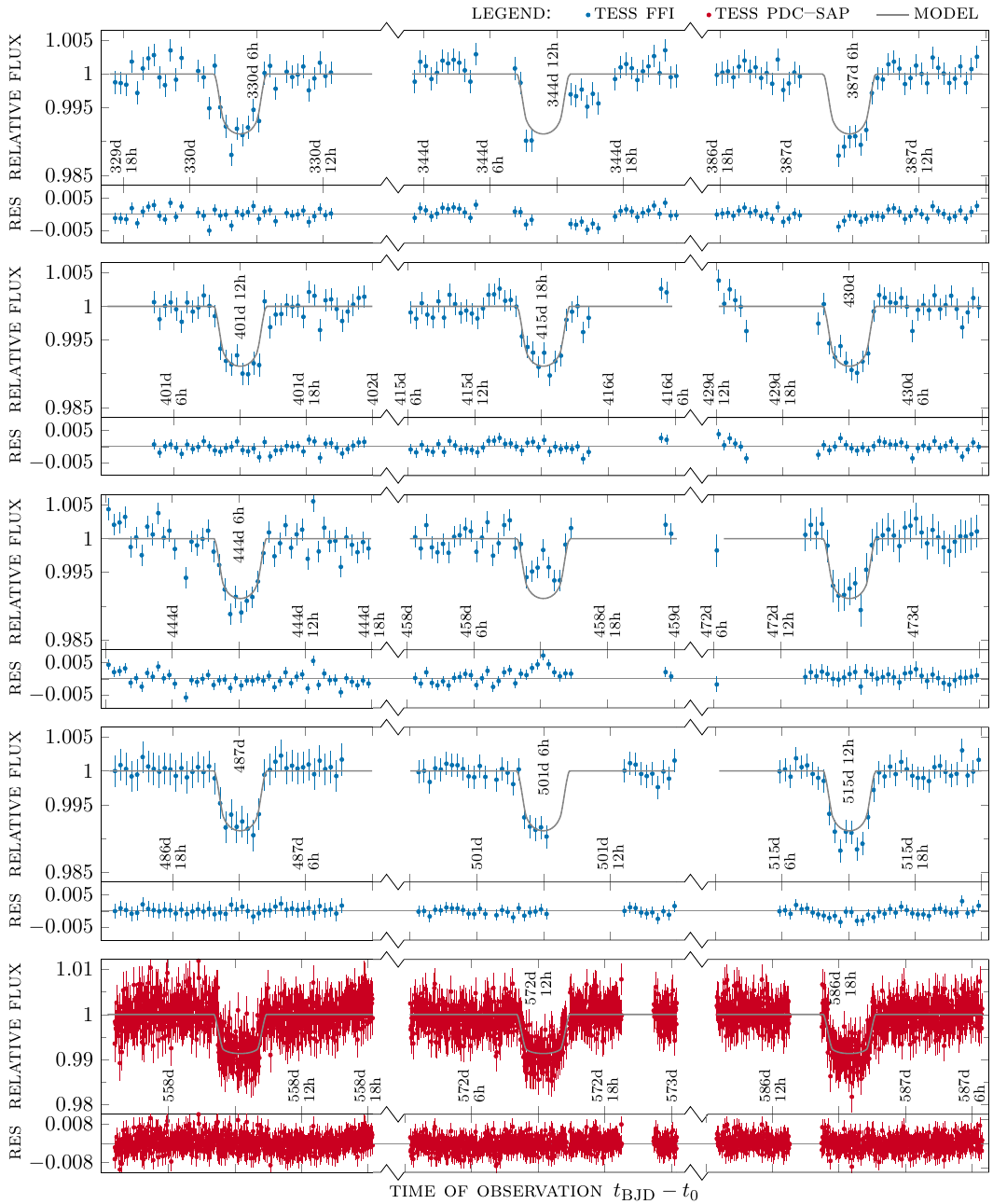}
        \subcaption{\tess\ 30-min \ffi\ photometry data of \toinum\ reduced with \tesseract\ (blue) and \tess\ 2-min cadence \pdc--\sap\ photometry data (red). The data and model (gray) shown are \(\pm \sfrac{1}{2}\) days from the fitted mid-transit time during \ttv\ extraction.}
        \label{fig:flux1}
    \end{subfigure}
    \caption{Transit photometry around the transit events of \toinumb.}
\end{figure*}%
\begin{figure*}[t!]\ContinuedFloat
    \centering
    \begin{subfigure}{\textwidth}
        \includegraphics[scale=1]{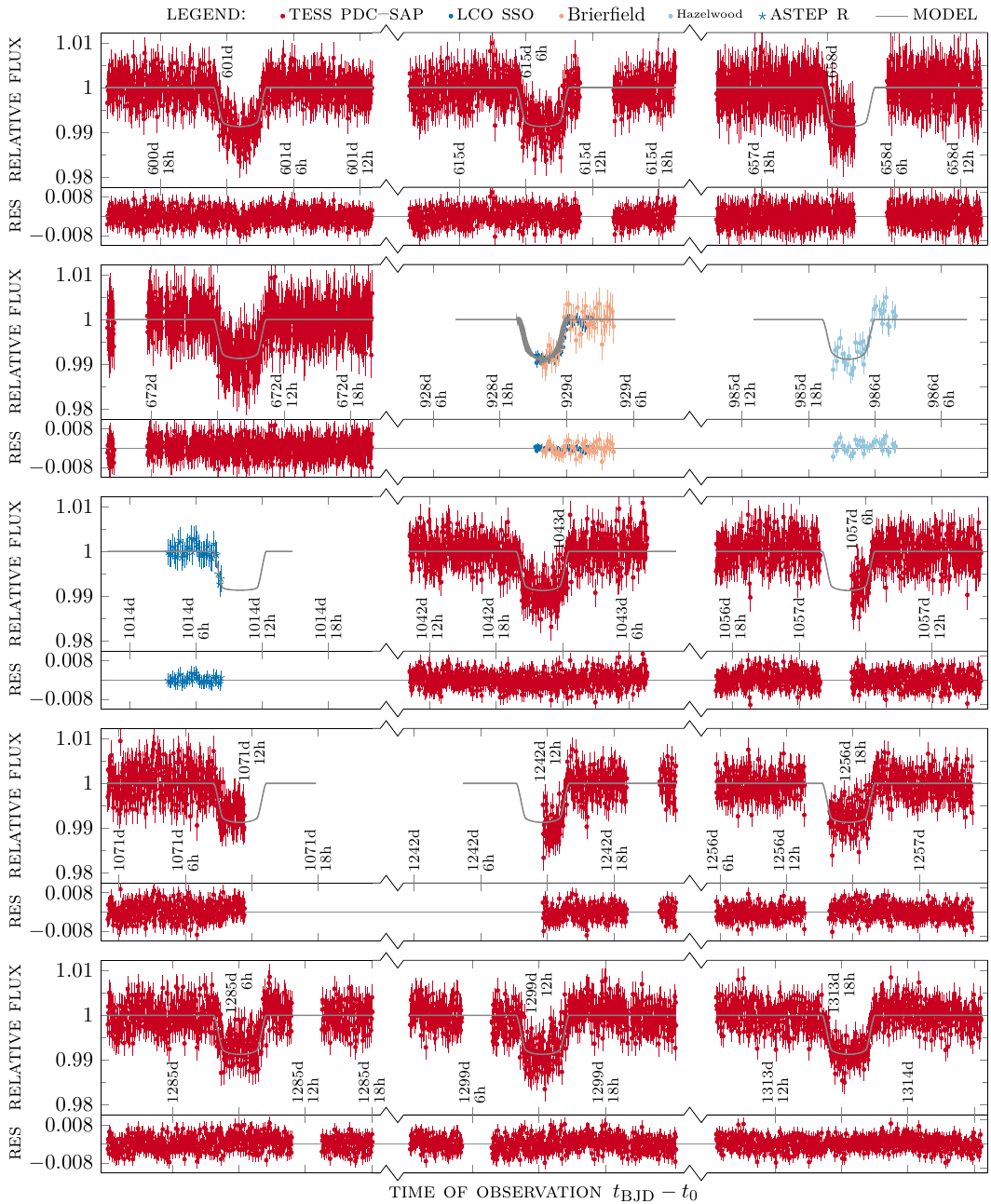}
        \subcaption{\tess\ 2-min cadence \pdc--\sap\ photometry data (red) and ground-based photometry (as indicated in the legend).
        The data and model (gray) are shown \(\pm \sfrac{1}{2}\) days from the fitted mid-transit time during \ttv\ extraction.}
        \label{fig:flux2}
    \end{subfigure}
    \caption{Transit photometry around the transit events of \toinumb.}
\end{figure*}
\begin{figure*}[t!]\ContinuedFloat
    \centering
    \begin{subfigure}{\textwidth}
        \includegraphics[scale=1]{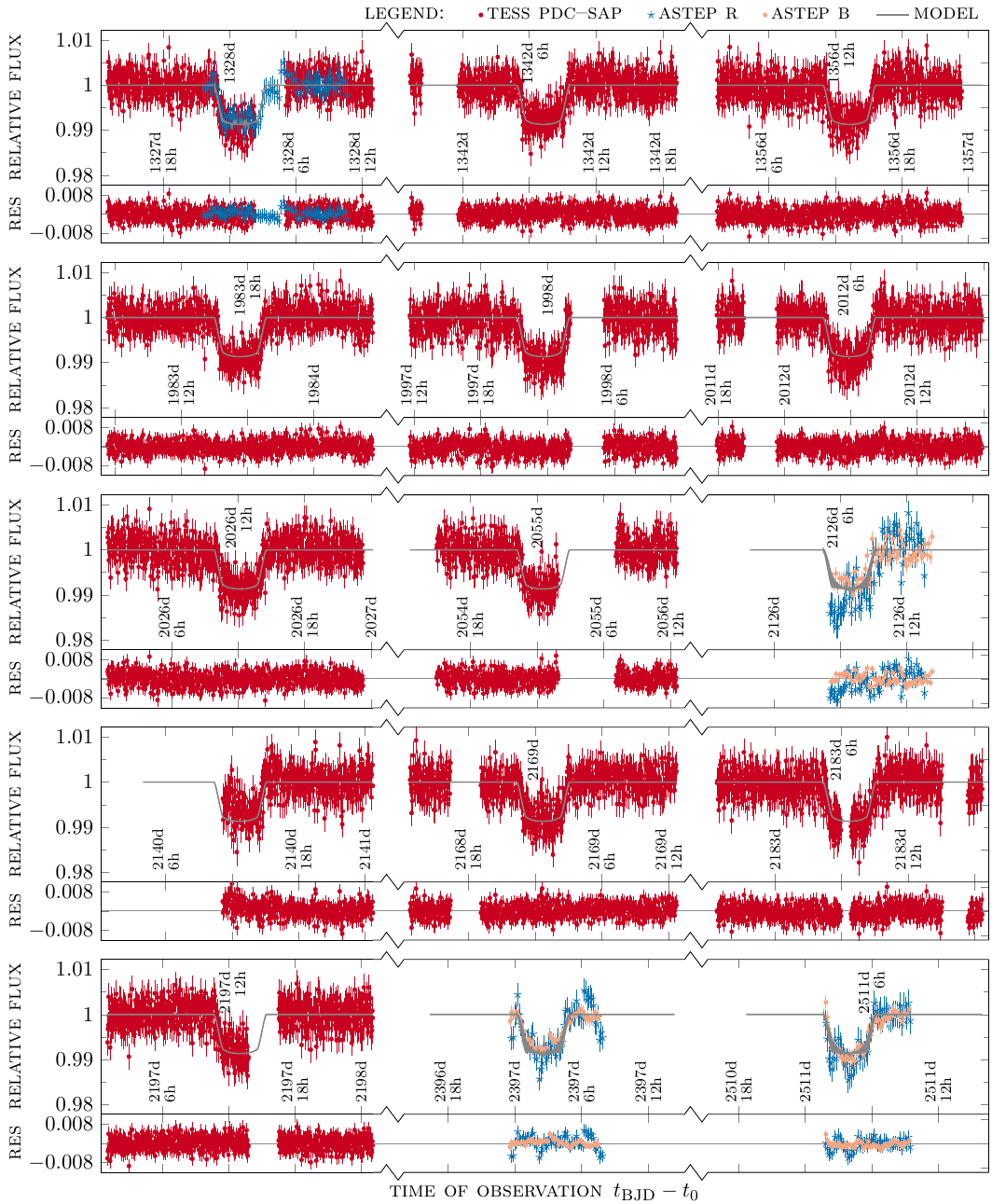}
        \subcaption{\tess\ 2-min cadence \pdc--\sap\ photometry data (red) and ground-based photometry (as indicated in the legend). 
        The data and model (gray) are shown \(\pm \sfrac{1}{2}\) days from the fitted mid-transit time during \ttv\ extraction.}
        \label{fig:flux3}
    \end{subfigure}
    \caption{Transit photometry around the transit events of \toinumb.}
    \label{fig:fluxes}
\end{figure*}

In this paper, we report the discovery of a warm Jovian-mass planet pair around the G-dwarf star \toinum. 
The inner planet, \toinumb, shows transit events with a period of \SI{14.33}{\,\days} detected by \tess. 
The transiting planet exhibits periodic transit timing variations (\ttv s), with a semi-amplitude of \(\approx\SI{27}{\min}\) and a \ttv\ ``super-period" of \(\approx\SI{236}{\,\days}\)\footnote{A \ttv\ ``super-period" quantifies the periodicity of the \ttv\ signal and can be approximated by:
\(P_{\rm{\ttv}} = {\left|(j-1) / P_{\rm{in}} - j / P_{\rm{out}}\right|}^{-1},\)
where \(j = P_{\rm{out}} / P_{\rm{in}}\) represents the close commensurability between the inner and the outer planets. In case of only one transiting planet, the ``super-period" could help determine the most likely period of a non-transiting exoplanet perturber \citep[see][and references therein]{Vitkova2025}.}, which, when combined with radial velocity (\rv) measurements taken with the \textsc{f}iber-fed \textsc{e}xtended \textsc{r}ange \textsc{o}ptical \textsc{s}pectrograph (\feros), revealed the existence of an additional outer Saturn-mass planet, \toinumc, with an orbital period of \(\sim \SI{30.4}{\,\days}\). 
This discovery was made in the context of our ongoing \textsc{w}arm g\textsc{i}a\textsc{n}ts with t\textsc{e}ss (\wine) survey, which aims at the systematic characterization of \tess\ transiting giant planets in warm orbits \citep[e.g.][]{hd1397, jordan2020, Brahm2020, Schlecker2020, Bozhilov2023}. 
We have detected a substantial sample of pairs of transiting warm giants with strong \ttv s \citep[e.g.,][]{Trifonov2021, Trifonov2023, Hobson2023, Vitkova2025}.

In \autoref{sec2}, we present the transit photometry and Doppler observational data used to reveal the existence of \toinumb\ and their orbital and physical parameters. 
We derive atmospheric and physical stellar parameter estimates of \toinum\ in \autoref{sec3}, which are based on our \feros\ spectra and \parsecabbr\ stellar evolution models.
In \autoref{sec4}, we describe our data and orbital analysis approach.
In \autoref{sec5}, we present an overview of the dynamical architecture and long-term stability properties of the \toinum\ system. 
In \autoref{sec6}, we discuss the possible internal composition of the transiting exoplanet \toinumb, and finally, in \autoref{sec7}, we present our summary and conclusions.

\section{Data}
\label{sec2}
\noindent
This work combines \tess\ photometry with ground-based transit observations, high-resolution Speckle imaging, and precision spectroscopy to characterize the \toinum\ system. 
The datasets and their respective reduction procedures are described in the following pages.

\subsection{TESS}
\label{Sec2.1}
\noindent
The system \toinum\ was observed by \tess\ in a total of 27 Sectors.
Full frame image (\ffi) data with \qty[number-unit-product=\text{-}]{30}{\min} cadence were taken in Sectors 1, 3, 4, 5, 6 and 7, whereas \qty[number-unit-product=\text{-}]{2}{\min} cadence data were taken in Sectors 9, 10, 11, 13, 27, 28, 30, 31, 33, 34, 35, 36, 37, 38, 61, 62, 63, 64, 67, 68, and 69.
The \ffi\ photometry was extracted using the \tesseract\footnote{\url{https://github.com/astrofelipe/tesseract}} pipeline (Rojas et al., in prep). The \tesseract\ package performs simple aperture photometry on the \ffi s, leveraging the \texttt{TESSCut}~\citep{TESSCut} and \lightkurve~\citep{lightkurve} packages. 
The extracted \ffi\ photometry was then corrected for dilution contamination from nearby stars. 
The dilution factor was estimated using the \(R_{P}\) \gaia\ \textsc{dr}3 fluxes of \toinum\ and all sources within the \tess\ aperture for each sector. 
We calibrated the \tess\ photometry by applying the dilution correction method described in Eq.~(6) of \citet{Espinoza2019} to each \ffi\ sector.

The \qty[number-unit-product=\text{-}]{2}{\min} cadence \tess\ data were retrieved from the Mikulski Archive for Space Telescopes\footnote{\url{https://mast.stsci.edu/portal/Mashup/Clients/Mast/Portal.html}} and can be retrieved from\dataset[10.17909/fwdt-2x66]{http://dx.doi.org/10.17909/fwdt-2x66} \citep{10.17909/fwdt-2x66}. 
For this target, the Science Processing Operations Center \citep[\spoc,][]{SPOC} at \nasa\ Ames Research Center provided both simple aperture photometry (\sap) and systematics-corrected photometry, derived using the \textit{Kepler} Presearch Data Conditioning (\pdc) algorithm \citep[\pdc,][]{Smith2012, Stumpe12}. 
The \pdc\ light curve was obtained by detrending the \sap\ light curve with a linear combination of co-trending basis vectors (\textsc{cbv}s), which were derived via principal component decomposition of light curves for each sector, camera, and \ccd. 
The \pdc-\sap\ light curves are further corrected for dilution from nearby field stars and instrumental systematics.

We suspected that the \qty[number-unit-product=\text{-}]{1.37}{day} signal comes from neighboring stars in the relatively large \tess\ aperture. 
We investigated if the transit signal was coming from neighboring stars by inspecting the target pixel file (\tpf) image data for each \tess\ sector. 
\autoref{TPFnew} shows the \tpf\ images of \toinum\ constructed from the \tess\ image frames and \gaia\ \textsc{dr}3 data \citep{Gaia_Collaboration_2021} of \toinum\ for Sectors 1, 3, and 4. 
The remainder of the \tpf\ images of \tess\ sectors for \toinum\ are shown in \autoref{fig:tpf2} and \autoref{fig:tpf3}.

The binary signal contaminating the \toinum\ photometry was identified as a planet candidate in the \ffi\ \tess\ light curves after following the \wine\ detection methods \citep[see][]{Brahm2020, Trifonov2020, Hobson2023}, finding a periodicity of the transits of \(\approx \qty{1.37}{\,\days}\). 
The signal has a typical transit shape which could be easily attributed to an eclipsing binary (\textsc{eb}). 
The raw \ffi\ and \qty[number-unit-product=\text{-}]{2}{\min} cadence light curves are visualized in \autoref{fig:relfluxes}, where a trained eye can detect the characteristic eclipsing binary V-shaped pattern signal in the time series.

We identified a background source within the \tess\ aperture, \qty{13}{\arcsec} away from \toinum, namely \ticnum\ (\gaianum, indicated as ``Source 2" in \autoref{TPFnew}, with a brightness of \(G = \qty{14.733 \pm 0.001}{mag}\)) as the most likely signal of the \textsc{eb}. 
We conducted follow-up ground-based observations of this object with the \qty{60}{\cm} telescope of the Observatoire Moana located in El Sauce observatory in Chile during the predicted transits of the \qty[number-unit-product=\text{-}]{1.37}{day} period. 
We detected the eclipsing binary signal as expected, firmly confirming that \ticnum\ is the \textsc{eb} contaminating the flux of \toinum. 
\autoref{fig_contaminator} shows the observational data of \ticnum\ which exhibits deep eclipses at the predicted times in three different epochs.  
The neighbor target \ticnum\ is an F-dwarf with an estimated stellar radius of \(\sim \SI{1.2}{\Rsun}\), thus if we neglect the flux from a secondary component of the \textsc{eb}, the observed \(\sim 10\%\) eclipse depth translates into a \(\sim \qty{0.4}{\Rsun}\) companion, consistent with an M3V red dwarf star.

Fitting the \qty[number-unit-product=\text{-}]{1.37}{day} signal with a transit model left residuals, revealing the transiting exoplanet \toinumb~with a period of \(\sim \qty{14.3}{\,\days}\).
To avoid having to use a complex model that involves the binary component to model the light curves, we used the \texttt{wotan} package \citep{Hippke2019} to identify the individual occultation events, mask them, and remove them from the light curves. 
The resulting detrended light curves are later used for our transit analysis and are shown in \autoref{fig:fluxes}.

Finally, the signature of the \qty[number-unit-product=\text{-}]{1.37}{day} period \textsc{eb} was detected by the \spoc\ in a search of Sector~9 with a noise-compensating matched filter \citep{Jenkins2002, Jenkins2010, JonJenkins2020}. 
The transit-like signature was fitted with an initial limb-darkened transit model \citep{Li2019} and subjected to a suite of diagnostic tests to help confirm or refute its planetary nature \citep{Twicken:DVdiagnostics2018}, including the difference image centroiding test, which located the source of the eclipses to the nearby star \ticnum\ located \qty{13}{\arcsec} from \toinum. 
A transit search of Sector~10 identified the \qty[number-unit-product=\text{-}]{14.2}{day} period signature of \toinumb\ and constrained the location of the host star to within \qty{0.768 \pm 2.6}{\arcs}.

\subsection{Ground--based transit photometry}
\label{ground}
\noindent
The \tess\ pixel scale is \(\sim \qty{21}{\!\arcsec \; \pixel^{-1}}\) and photometric apertures typically extend out to roughly \(\sim \qty{1}{\!\arcmin}\), generally causing multiple stars to blend in the \tess\ photometric aperture. 
To attempt to determine the true source of the \tess\ detection, we acquired ground-based time-series follow-up photometry of the field around \toinum\ as part of the \tess\ Follow-up Observing Program \citep[\textsc{tfop},][]{collins:2019}\footnote{\href{https://tess.mit.edu/followup}{tess.mit.edu/followup}}. 
We used the \texttt{TESS Transit Finder}, which is a customized version of the \texttt{Tapir} software package \citep{Jensen:2013}, to schedule our transit observations. 
A \(\sim \qty{9}{ppt}\) event was detected on-target in all observations. 
The light curve data are available on the \texttt{EXOFOP-TESS} website\footnote{\href{https://exofop.ipac.caltech.edu/tess/target.php?id=364395234}{exofop.ipac.caltech.edu/tess/target.php?id=364395234}} and are included in the transit modeling and \ttv\ extraction described in section \autoref{sec4}.

\subsubsection{Brierfield Observatory}
\noindent
We observed a partial transit event (egress) on \utc\ 2020 March 20 in the \(B\)-band from Brierfield Observatory near Bellingen, New S. Wales, Australia. 
The 0.36\,m telescope is equipped with a \(4096 \times 4096\) Moravian 16803 camera. 
The image scale after binning \(2 \times 2\) is \(1\farcs47\,\si{\pixel^{-1}}\), resulting in a \(\qty{50}{\arcmin} \times \qty{50}{\arcmin}\) field of view. 
The differential photometric data were extracted using \texttt{AstroImageJ} (\texttt{AIJ}) \citep{Collins:2017} with a circular \(4\,\farcs4\) photometric aperture that excluded all of the flux from the nearest known neighbor in the \gaia\ \textsc{dr}3 catalog (5289147741356513024), which is \(6\,\farcs 9\) northeast of \toinum. 
This object is more than 8 magnitudes fainter, so could not possibly be the source of the \tess\ transit detection, and furthermore it contributes negligible dilution in any photometric aperture.

\subsubsection{LCOGT-SSO}
\noindent
We observed a partial transit event (egress) on \utc\ 2020 March 20 in Johnson/Cousins \(I\)-band from the Las Cumbres Observatory Global Telescope (\lcogt) \citep{Brown:2013} \qty{1}{\meter} network node at Siding Spring Observatory (\sso) near Coonabarabran, Australia. 
The \qty{1}{\meter} telescope is equipped with a \(4096 \times 4096\) Sinistro camera having an image scale of \(0\,\farcs 389\) per pixel, resulting in a \(\qty{26}{\arcmin} \times \qty{26}{\arcmin}\) field of view. 
The images were calibrated by the standard \lcogt\ \texttt{BANZAI} pipeline \citep{McCully:2018} and differential photometric data were extracted using \texttt{AIJ} with a circular \(3\,\farcs 5\) photometric aperture.

\begin{figure}
    \centering
    \includegraphics[scale=1]{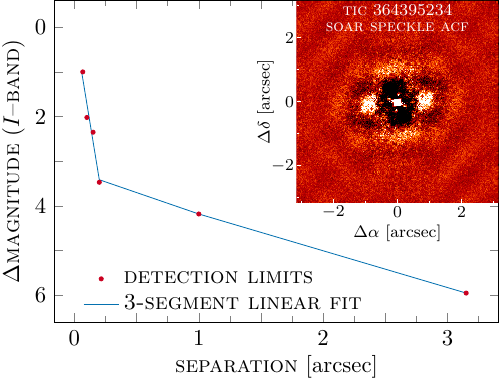}
    \caption{\soar\ speckle observations of \toinum. The curve represents the \(5\sigma\) detection sensitivity, and the inset shows the speckle imaging autocorrelation function (\textsc{acf}). No companion was detected within the contrast limit.}
    \label{fig:soar}
\end{figure}

\subsubsection{Hazelwood Observatory}
\noindent
We observed a partial transit event (egress) on \utc\ 2020 May 16 in Johnson/Cousins \(R\)-band from Hazelwood Observatory near Churchill, Victoria, Australia. 
The \qty{0.32}{\meter} telescope is equipped with a \(2184 \times 1472\) \textsc{sbig} \textsc{stt}3200 camera. 
The image scale is \(0\,\farcs 55\,\si{\pixel^{-1}}\), resulting in a \(\qty{20}{\arcmin} \times \qty{14}{\arcmin}\) field of view. 
The photometric data were extracted using \texttt{AIJ} with a circular \(6\,\farcs 6\) photometric aperture.

\subsubsection{ASTEP}
\noindent
Between 2020 and 2024, we observed five transit windows of \toinumb\ with the Antarctica Search for Transiting ExoPlanets (\astep) program on the East Antarctic plateau \citep[][]{Guillot15, Mekarnia16}. 

Until December 2021 \astep\ was equipped with a \(\qty{4}{\kcount} \times \qty{4}{\kcount}\) front-illuminated \textsc{fli} \textsc{p}roline \textsc{kaf}--16801\textsc{e} \ccd\ with an image scale of \(0\,\farcs 93\,\si{\pixel^{-1}}\), resulting in a \(\qty{1}{\degree} \times \qty{1}{\degree}\) corrected field of view. 
The focal instrument dichroic plate splits the beam into a blue wavelength channel for guiding, and a non-filtered red science channel roughly matching an \(R_{\mathrm{c}}\) transmission curve \citep{Abe13}. 
In January 2022, the focal box was replaced with a new one with two high sensitivity cameras including an Andor iKon-L 936 at red wavelengths with a transmission curve centered on \qty{850 \pm 138}{\nano\meter} and an image scale of \(1\farcs 30\,\si{\pixel^{-1}}\), and an \textsc{fli} \textsc{kl}400 at ``blue" wavelengths centered on \qty{550 \pm 150}{\nano\meter} with a scale of \(1\farcs 06\,\si{\pixel^{-1}}\) \citep{Schmider22}.

Because of the low data transmission rate at the Concordia Station, the photometry time series are processed on-site using Interactive Data Language (\textsc{idl}) \citep{IDL} and Python aperture photometry pipelines \citep{Mekarnia16, Dransfield22}.

The first observations in the old setting, on June 13, 2020 and April 23, 2021, showed an ingress and a full \(\sim 6 \text{ to } 8\,\,\si{ppt}\) transit occurring later than predicted by about 15 and 50 minutes, respectively. 
Three observations in the new setting took place on June 30, 2023, March 27, 2024, and July 19, 2024, leading to the detection of an egress and two full transits, respectively, all occurring 70 to 110 minutes earlier than the linear prediction at the time. 
The depths of these events, between 7 and 8.5 ppt on both cameras, were consistent with achromatic transits. 
All five observations occurred with temperatures between \SI{-70}{\celsius} and \SI{-60}{\celsius} and winds between 3 and \SI{6}{\m\per\s}, clear conditions except the 2023 egress that was affected by high clouds. 
The aperture radius, chosen to maximize photometric quality, was between \(7\,\farcs 8\) and \(11\,\farcs 3\).

\begin{table}[bp]
\caption{Stellar parameters of \toinum\ (\ticnum,
Gaia \textsc{dr}3\,5289147737059634560) and their \(1 \sigma\) uncertainties derived using \texttt{zaspe} spectral analyses, \gaia\ parallax, broadband photometry, together with advanced \parsecabbr\ models.}
\label{table:phys_param}
\begin{tabularx}{\columnwidth}{p{3.8cm} l r}
\toprule
  \textsc{parameter [units]} & \toinum & \textsc{reference} \\[2pt]
\midrule
   Spectral type & G1V & [1] \\
   \(B\) magnitude & \(13.397 \pm 0.489\) & [2] \\
    \(V\) magnitude & \(12.68 \pm 0.23\) & [2] \\
   \(G\) magnitude & \(12.258 \pm 0.003\) & [3] \\
   \(K\) magnitude & \(10.70\pm 0.02\) & [4] \\
   Distance [\unit{\parsec}] & \(1213_{-68}^{+74}\) & [3] \\
   Mass [\(M_{\odot}\)] & \(1.058_{-0.063}^{+0.065}\)  & \thispap \\
   Radius [\(R_{\odot}\)] & \(1.072 \pm 0.047\) & \thispap \\
   Luminosity [\(L{_\odot}\)] & \(1.183_{-0.060}^{+0.063}\) & \thispap \\
   Age [\unit{\giga\year}] & \(4.0_{-1.9}^{+2.0}\) & \thispap\\
   \(A_{V}\) [\unit{\mag}] & \(0.307 \pm 0.067\) & \thispap \\
   \(T_{\mathrm{eff}}\) [\unit{\kelvin}] & \(5823 \pm 161\) & \thispap \\
   \(\log(g) \left[\unit{\centi\metre\cdot\second^{-2}}\right]\) & \(4.403 \pm 0.02\) & \thispap \\
   \([\rm{Fe/H}]\) & \(0.18 \pm 0.05\) & \thispap \\
   \(v \sin(i) \left[\unit{\kilo\metre\cdot\second^{-1}}\right]\) & \(4.13 \pm 0.3\) & \thispap \\
\bottomrule
\end{tabularx}

\tablecomments{\small [1] \citet{ESA}, [2] \citet{2000A&A...355L..27H}, [3] \citet{Gaia_Collaboration2016, Gaia_Collaboration2018b,Gaia_DR3_2023}, [4] \citet{Cutri2003}.}
\end{table}

\subsection{Speckle imaging}
\noindent
We inspected the speckle imaging frame obtained with the \qty{4.1}{\meter} \soar\ telescope \citep{SOAR} as part of the \soar\ \tess\ survey \citep{SOAR_TESS}. 
Observations of \toinum\ (\textsc{tic}\,364395234) were conducted on 9 November 2019, using the Cousins \(I\)-filter with \textsc{hrc}am. 
\autoref{fig:soar} shows the speckle auto-correlation function along with a contrast curve, achieving a contrast of \(\Delta \rm{mag} = 6\) at \SI{1.0}{\arcs} and an estimated point spread function of \SI{0.067}{\arcs}. 
No apparent contaminants were detected within \(\sim 3 \arcsec\) of the target star, ruling out the possibility of contamination from nearby sources.

\subsection{FEROS}
\label{FEROS}
\noindent
\toinum\ was a primary target in the \wine\ survey for spectroscopic follow-up using the \feros\ spectrograph on the \textsc{mpg}\footnote{Max-Planck-Gesellschaft}/\textsc{eso} 2.2\,m telescope at La Silla Observatory \citep{Kaufer1999}. 
An observational campaign on \toinum\ was conducted with \feros\ from March 2, 2020, to June 7, 2023, yielding 21 high-quality spectra obtained via the ThAr simultaneous calibration method. 
Exposure times ranged from 1200 to 1500 seconds, providing an average signal-to-noise ratio of 60 per resolution element. 
The \feros\ spectra were calibrated, and \rv s were extracted with the \texttt{ceres} pipeline \citep{ceres}, following procedures applied in other \wine\ survey studies \citep[e.g.,][and references therein]{Espinoza2020, Brahm2020, Hobson2023, Eberhardt2023}. 
In addition to precise \rv s, we derived bisector span and full width at half maximum (\fwhm) measurements as stellar activity diagnostics. 
The median \rv\ uncertainty for \toinum\ is approximately \SI{9}{\m\per\s}, with an \textsc{rms} of \SI{80.6}{\m\per\s}. 
The \rv s and spectral line stellar activity measurements are listed in \autoref{tab:FEROS_RVS}.

\section{Stellar parameters of \toinum}
\label{sec3}
\noindent
The stellar atmospheric parameters of \toinum\ were derived from the co-added \feros\ spectra using the \texttt{zaspe} package \citep{zaspe}, the same way as was done for other \wine--\tess\ exoplanet systems \citep[see e.g.,][for more details.]{Brahm2020, Hobson2023}. 
The \texttt{zaspe} tool provides \(T_{\rm{eff}}\), \(\log(g)\), \(v \sin(i)\), and [Fe/H] comparing the obtained spectra to a grid of synthetic spectra generated from \textsc{atlas}\,9 model atmospheres \citep{atlas9}.

The fundamental parameters of \toinum\ were derived using the \parsecabbr\ evolutionary models, following the approach outlined in \citet{hd1397}. 
These models allow us to compare the absolute magnitudes corresponding to a given set of stellar parameters to those of the target star, incorporating the distance to \toinum\ from the \gaia\ \textsc{dr}3 catalog \citep{Gaia_Collaboration_2021}. 
For this comparison, we utilized the \gaia\ \(G\), \(G_{\mathrm{BP}}\), and \(G_{\mathrm{RP}}\) magnitudes, along with the 2\textsc{mass} \(J\), \(H\), and \(K_{\mathrm{s}}\) bands. 
The metallicity was fixed to the value obtained from \texttt{zaspe}, and the stellar age and \(M_{\star}\) parameter space was explored using the \texttt{emcee} package \citep{emcee}. 
This analysis yielded a more precise estimate of \(\log(g)\) than the initial spectroscopic determination. 
To refine this estimate, we employed an iterative procedure in which the \(\log(g)\) value from the \textsc{sed} analysis was used as input for a subsequent \texttt{zaspe} run, with iterations continuing until convergence was reached.

\begin{figure}[t]
    \centering
    \includegraphics[scale=1]{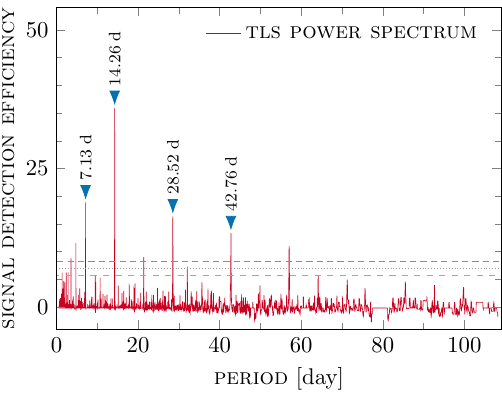} 
\caption{
\tls\ periodogram of the detrended and binary corrected \tess\ light curve data for \toinum. 
The planetary transit signal at \(P_{\rm b} = \qty{14.26}{d}\) is accompanied by harmonics at \qty{7.13}{d}, \qty{28.52}{d}, and others. 
Horizontal dashed lines represent the signal detection efficiency \citep[\textsc{sde},][]{Hippke2019} thresholds of 5.7, 7.0, and 8.3, corresponding to \tls\ false positive rates of 10\%, 1\%, and 0.1\%, respectively.}
\label{TLS_results} 
\end{figure}

The atmospheric and physical parameters are listed in \autoref{table:phys_param}. 
We found that \toinum\ is consistent with a G-type dwarf star with a mass of \(\qty[parse-numbers=false]{1.058_{-0.035}^{+0.037}}{\Msun}\) and a radius of \(\qty{1.072 \pm 0.012}{\Rsun}\), exhibiting slightly higher metallicity than solar. 
We note that typically, the obtained uncertainties in the stellar parameters from our analysis are relatively small. 
Our estimates do not account for well-known systematic differences relative to other stellar models. 
Therefore, we typically inflated the stellar parameter uncertainties to more realistic values following the prescription of \citet{Tayar2020}, which are adopted throughout this work and are also listed in \autoref{table:phys_param}.

\section{Analysis and Results}
\label{sec4}
\noindent
In this section we present the analysis of the photometric and spectroscopic datasets and derive the system parameters of \toinum. 
We first describe the modeling framework and software tools adopted to search for periodic signals and to perform transit and radial-velocity fits. 
We then extract transit times from the \tess\ and ground-based light curves, assess the Doppler content and activity diagnostics of the \feros\ spectra, and finally perform a joint \ttv+\rv\ dynamical analysis to constrain the architecture of the multi-planet system.

\subsection{Tools}
\label{Sec4.1}
\noindent
For our data analysis, we employed the publicly available \texttt{Exo-Striker} exoplanet toolbox\footnote{\url{https://github.com/3fon3fonov/exostriker}} \citep{Trifonov2019_es}. 
We performed detailed orbital and statistical analysis of the \rv\ and transit data by utilizing the generalized Lomb--Scargle periodogram \citep[\gls,][]{Zechmeister2009} and the \texttt{transitleastsquares} package \citep[\tls,][]{Hippke2019b}, respectively. 
These tools are incorporated into the graphical user interface of \texttt{Exo-Striker}, enabling efficient analysis of \rv\ and transit photometry signals. 
For transit light curve models, the \texttt{Exo-Striker} utilizes the \texttt{BAsic Transit Model cAlculatioN} package \citep[\texttt{batman};][]{Kreidberg2015}.

The \texttt{Exo-Striker} can fit exoplanet data using a self-sufficient \(N\)-body dynamical model, which is suitable for systems with gravitationally interacting planets like \toinum. 
Dynamical modeling of \rv s is internal to the \texttt{Exo-Striker}, whereas \ttv s are modeled using a Python wrapper of the \ttv \textsc{f}ast package \citep{Deck2014}. 
Orbital analysis was performed adopting a joint-model approach by fitting the \feros\ \rv s and the extracted \ttv s, where the input and output parameters are expressed as Jacobi elements \citep[e.g.,][]{Lee2003}.

Our orbital modeling scheme includes a combination of three parameter optimization schemes for efficient derivation of the parameter posteriors and best-fit models. 
To scan the parameter space more efficiently, we rely on detailed Bayesian analysis employing the Nested Sampling (\ns) scheme \citep{Skilling2004} via the \texttt{dynesty} sampler \citep{Speagle2020}, followed by parameter optimisation via the Simplex (Nelder-Mead) method, which allows us to pinpoint the best-fit solution in terms of maximum \(\ln\mathcal{L}\). 
Finally, the posteriors around the best-fit were accessed using the affine-invariant ensemble Markov Chain Monte Carlo (\mcmc) sampler \citep{Goodman2010} through the \texttt{emcee} package \citep{emcee}.

\begin{table}[t]
    \caption{Planetary radii, orbital inclination for planet \toinumb, derived during \ttv\ extraction.}
    \label{table:table2}
    \begin{tabularx}{\columnwidth}{l@{\hskip 100pt} r}
    \toprule
    \textsc{parameter [units]} & \textsc{median} \\[2pt]
    \midrule
    \(a_{b} / R_{\star}\) & \(20.0_{-0.6}^{+0.8}\) \\
    \(r_{b} / R_{\star}\) & \(0.0914_{-0.0001}^{+0.0001}\) \\
    \(i_{b}\) [deg] & \(88.05_{-0.13}^{+0.17}\) \\
    \(r_b\) [\(R_{\mathrm{Jup}}\)] & \(0.975_{-0.052}^{+0.053}\) \\
    \bottomrule
    \end{tabularx}
\tablecomments{The remaining transit light curve data parameter estimates are listed in \autoref{tab:tableA1}, whereas the individual transit times for \toinumb\ are listed in \autoref{table:TTVdataUpdated}.}
\end{table}

\subsection{Transit light curve analysis}
\label{Sec4.2}
\noindent
As first steps in our semi-automated \wine\ pipeline of \tess~data analysis, we inspected the \ffi\ light curves of \toinum\ obtained using \tesseract. 
We masked the eclipsing binary signal (see \autoref{Sec2.1}), and we fit each of the \texttt{tesseract} \ffi\ sector light curves with a robust (iterative) Mat\'ern Gaussian process (\textsc{gp}) kernel. 
The \textsc{gp} model was applied to effectively capture and de-trend any systematic variation of the light curves \citep[see][]{Hippke2019}. 
We scan the combined \ffi\ light curve for transit events by adopting the \texttt{transitleastsquares} package \cite[\tls,][]{Hippke2019b}. 
\autoref{TLS_results} shows the constructed \tls\ power spectra of the available \tess-\ffi\ transit data of \toinum. 
We detected a significant power peak with a period of \(P_{\rm b} = \qty{14.26}{\,\days}\). 
The remaining periodic signals detected in the \tls\ spectra, e.g., those peaked at \qty{7.13}{\,\days}, at \qty{28.52}{\,\days}, etc., are unavoidable harmonics of the transit signal.

\begin{figure}[t!]
    \begin{subfigure}{0.47\textwidth}
        \includegraphics[scale=1]{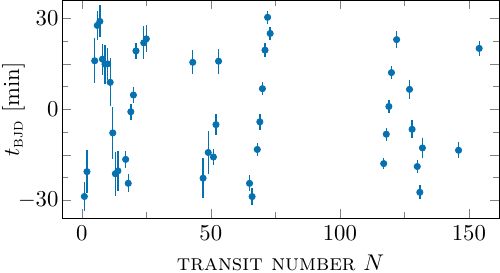}
        \subcaption{The extracted \ttv s (and their associated errors) of \toinumb.}
        \label{fig:gls1}
    \end{subfigure}
    \begin{subfigure}{0.47\textwidth}
        \includegraphics[scale=1]{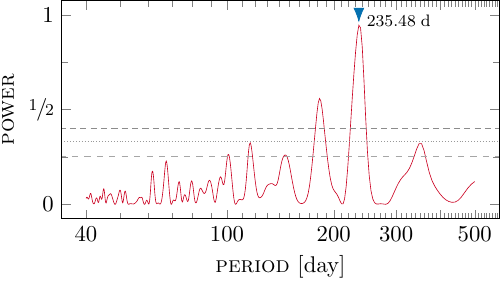}
        \subcaption{\gls\ periodogram of the \ttv\ measurements containing a significant period (\ttv\ super-period) of 235.5 days.}
        \label{fig:gls2}
    \end{subfigure}
    \caption{Transit photometry around the transit events of \toinumb.}
\label{TTV_gls}
\end{figure}

\begin{figure}[tp]
\centering
\includegraphics[scale=1]{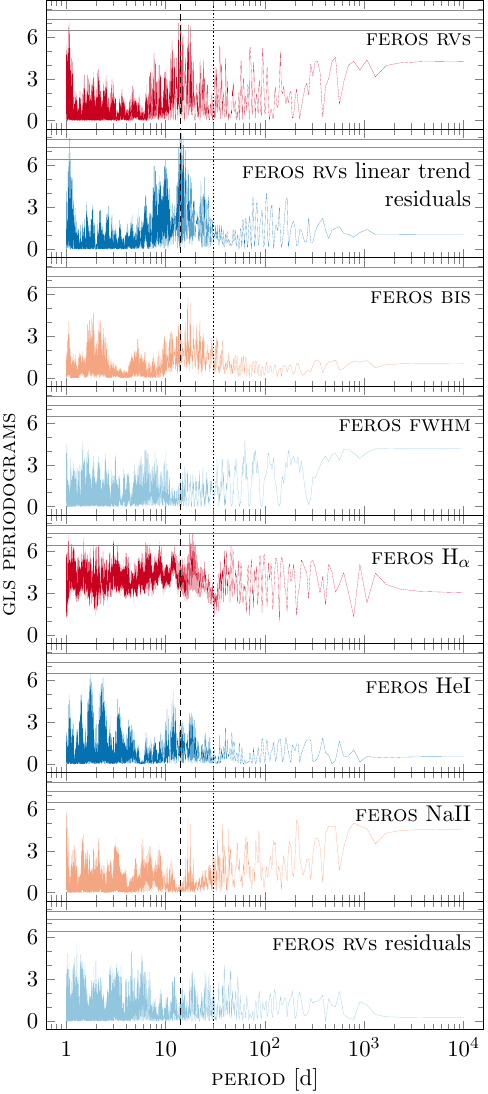}
\caption{\gls\ power spectrum for the \toinum\ data, based on \feros\ \rv s spectra. 
     From top to bottom panels, as labeled, \rv s used in this work, \rv\ residuals 
     of the linear trend model being applied to the \rv s, \bis, \fwhm, 
     $\rm{H}_{\alpha}$, He\,I,  Na\,II, and the residuals of our final best fit \ttv+\rv\ self-consistent dynamical model, respectively. The dashed and dotted vertical lines indicate the orbital period of \toinumb\ and \toinumc, respectively.
     The horizontal lines in the \gls\ periodograms show the FAP levels 
     of 10\%, 1\%, and 0.1\%.}
\label{MLP_results} 
\end{figure}

\begin{figure*}[tp]
    \includegraphics[scale=1]{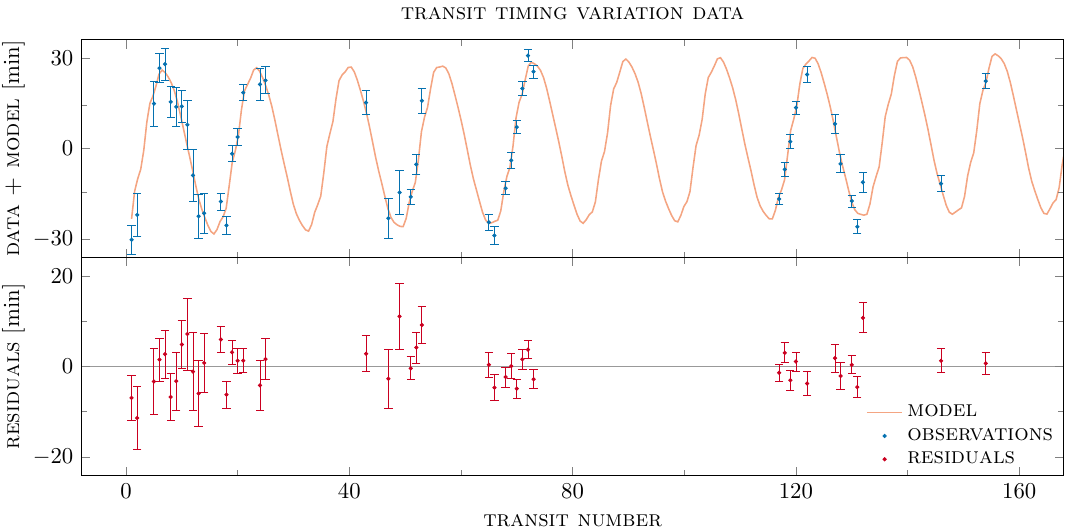}
\caption{\ttv\ measurements of \toinumb\ modeled simultaneously with the \rv s of \toinum\ from \feros\ using a two-planet dynamical model. 
\emph{Top panel}: Presents the \ttv\ time series around a 
mean (osculating) period of P = 14.25503 days,
alongside a model consistent with two massive planets near the 2:1 \mmr. 
\emph{Bottom panel}: Displays the residuals of the \ttv s after subtracting the best-fit model.}
\label{TTV_plot1}
\end{figure*}

Fitting a single-planet model to the \ffi\ data using the \texttt{Exo-Striker}, however, led to an obvious mismatch of the observed transits of \toinumb. 
We optimized the transit light curve model by fitting \toinumb's orbital period \(P_{\rm b}\), the eccentricity \(e_{\rm b}\) and  argument of periastron \(\omega_{\rm b}\) via the parametrisation \((e_{\rm b} \sin\omega_{\rm b}, e_{\rm b} \cos\omega_{\rm b})\), the inclination \(i_{\rm b}\), the time of inferior transit conjunction \(t_{0}\), the semi-major axis relative to the stellar radius \(a_{\rm b} / R_{\star}\), and the planetary radius relative to the stellar radius \(r_{\rm b} / R_{\star}\) (see \autoref{table:table2}). 
The \tess-\ffi\ light curve transit signals showed strong deviations in the expected individual times-of-transit from the derived posteriors of the exoplanet period, suggesting notable \ttv s. 
Thus, we followed the \toinum~over the \tess~mission cycles and collected convincing evidence that the \ttv s are present. 
Meanwhile, more ground-based photometric observations of \toinum~(see \autoref{sec2}), solidify our findings and allow us to reveal the physical and orbital parameters of the \toinum~system.

\begin{figure*}[!ht]
    \includegraphics[scale=1]{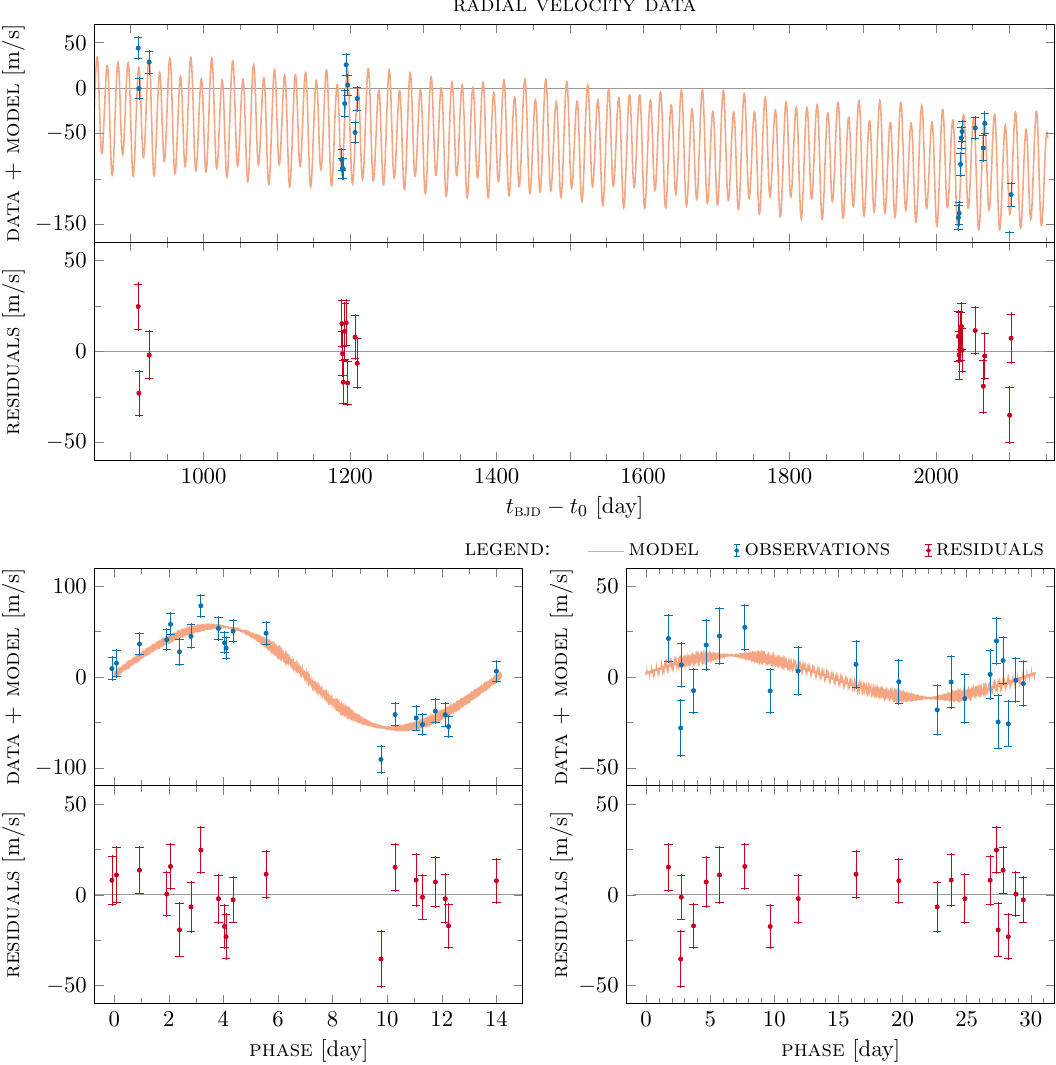}
    \caption{The top panel shows the \rv s of \toinum\ from the \feros\ spectrograph together with the \rv\ component of the \rv+\ttv\ joint \(N\)-body model. 
The model is consistent with two massive planets with an additional \rv\ linear trend component, suggesting a likely third sub-stellar body in the system. 
The bottom sub-panel shows the \rv\ residuals after the model contribution has been subtracted. 
The bottom left and right panels show a phase-folded representation of the model and \rv\ data.}
\label{TTV_plot2} 
\end{figure*}

To extract the \ttv s, we performed a separate one-planet fit to the full set of \tess\ and ground-based light curves, assuming a circular orbit (i.e., \(e_{\rm b} \sin(\omega_{\rm b}) = 0, e_{\rm b} \cos(\omega_{b}) = 0\)). 
In this numerical setup, we fit each recorded transit with its own parameter for the mid-transit times \(t_{n}\), where \(n\) is the transit number with respect to the first observed transit of \(\toinumb\).
The rest of the transit parameters in this modeling scheme were shared across each individual light curves, except for the limb-darkening (\ld) coefficients of \tess\ and the Brierfield, \lcogt-\sso, Hazelwood, and \astep, for which we adopted separate quadratic \ld\ set of parameters (to reflect the different filters used), and nuisance parameters such as the transit light curve relative photometric offset and additional photometric jitter which is added in quadrature to the error budget of the data sets \citep[see,][]{Baluev2009}. 
To perform an adequate parameter search, we ran a nested sampling scheme with 1000 ``live-points", focused on the posterior convergence instead of Bayesian evidence \citep[see,][for details]{Speagle2020}. 
We adopted the 68.3\% confidence levels of the nested sampling posterior probability parameter distribution as the \(1\sigma\) parameter uncertainties.

The extracted mid-transit time estimates and their precise uncertainties yielded strong periodic \ttv s in the photometric data. 
The top panel of \autoref{TTV_gls} shows the extracted \ttv s with respect to the mean Period of \toinumb, while the bottom panel of \autoref{TTV_gls} shows a \gls\ periodogram of the \ttv s. 
We find a significant \ttv\ semi-amplitude around the estimated mean orbital period of \(\sim \qty{27}{minutes}\), covering 10 full \ttv\ "super-periods" with a period of \qty{235.5(7)}{\,\days}. 
\autoref{table:TTVdataUpdated} lists the precisely extracted individual transit times and their errors. 
No significant \tls\ power is detected in the \tess\ photometry residuals, suggesting that only one planetary companion of \toinum\ is detectable in the available \tess\ light curves. 
These results reveal an additional non-transiting companion in the \toinum\ systems that is close and massive enough to perturb the Jovian-like transiting planet \toinumb.

\begin{table*}[ht!]
\caption{Nested sampling priors, posteriors, and the optimum \(-\ln\mathcal{L}\) orbital parameters of the two-planet system derived by joint \(N\)--body modeling of \ttv s (\tess) and \rv s (\feros).}
\label{table:NS_params}
\begin{tabularx}{\textwidth}{l @{\hskip 21pt} r r @{\hskip 21pt} r r @{\hskip 21pt} r r}
\toprule 
\multirow{2}{*}[-4pt]{\textsc{parameter [units]}} & \multicolumn{2}{c}{\textsc{median and} $1\sigma$} & \multicolumn{2}{c}{\textsc{maximum} $-\ln\mathcal{L}$} & \multicolumn{2}{c}{\textsc{adopted priors}} \\ [2pt] \cmidrule{2-7}
& \textsc{planet} b & \textsc{planet} c & \textsc{planet} b & \textsc{planet} c & \textsc{planet} b & \textsc{planet} c \\ [2pt]
\midrule
\(K \left[\si[per-mode=power]{\m\per\s}\right]\) & \(56.10_{-2.90}^{+2.84}\) & \(11.61_{-1.70}^{+1.26}\) & \(56.10\) &  \(11.78\) & \(\mathcal{U}\,(40.0, 100.0)\)  & \(\mathcal{U}\,(10.0, 50.0)\) \\
\(P \, \left[\si{\day}\right]\) & \(14.256_{-0.001}^{+0.001}\) & \(30.356_{-0.012}^{+0.010}\) & \(14.255\) & \(30.362\) & \(\mathcal{U}\,(14.0, 15.0)\) & \(\mathcal{U}\,(30.0, 31.0)\) \\
\(e \, \sin(\omega)\) & \(0.063_{-0.013}^{+0.020}\) & \(0.013_{-0.022}^{+0.035}\) & \(0.055\) & \(0.000\) & \(\mathcal{U}\,(-0.1, 0.1)\) & \(\mathcal{U}\,(-0.1, 0.1)\) \\
\(e \, \cos(\omega)\) & \(-0.007_{-0.015}^{+0.008}\) & \(0.003_{-0.021}^{+0.016}\) & \(-0.007\) & \(0.006\) & \(\mathcal{U}\,(-0.1, 0.1)\) & \(\mathcal{U}\,(-0.1, 0.1)\) \\
\(\lambda \, \si{\left[\deg\right]}\) & \(85.9_{-0.9}^{+1.6}\) & \(201.8_{-2.5}^{+3.4}\) & \(85.8\) & \(201.8\) & \(\mathcal{U}\,(0, 360)\) & \(\mathcal{U}\,(180, 200)\) \\
\(\incl \, \si{\left[\deg\right]}\) & \(89.44_{-0.48}^{+0.31}\) & \(87.99_{-0.46}^{+0.51}\) & \(89.48\) & \(87.99\) & \(\mathcal{N}\,(89.5, 0.5)\) & \(\mathcal{N}\,(88.0, 1.0)\) \\
\(\Omega \, \si{\left[\deg\right]}\) & \(12.24_{-3.30}^{+4.14}\) & \dots & \(12.22\) & \dots & \(\mathcal{N}\,(0.0, 30.0)\) & \textsc{fixed} \\
\(\mathrm{RV}_{\mathrm{off. \, FEROS}} \, \left[\si{\m\per\s}\right]\) & \(9616.26_{-4.71}^{+5.26}\) & \dots  & \(9616.26\) & \dots & \(\mathcal{U}\,(9500.0, 9700.0)\) & \dots \\
\(\mathrm{RV}_{\mathrm{jit. \, FEROS}} \, \left[\si{\m\per\s}\right]\) & \(9.90_{-1.76}^{+3.85}\) & \dots &  \(9.62\) & \dots & \(\mathcal{U}\,(1.0, 50.0)\) & \dots \\
\(\mathrm{RV}_{\mathrm{lin. tr. \, FEROS}} \, \left[\si{\m\per\s\per\day}\right]\) & \(-0.051_{-0.005}^{+0.005}\) & \dots & \(-0.051\) & \dots & \(\mathcal{U}\,(-0.1, 0.1)\) & \dots \\
\(e\) & \(0.065_{-0.013}^{+0.020}\) & \(0.027_{-0.016}^{+0.027}\) & 0.055 & 0.006 & \derived & \derived \\
\(\omega \, \si{\left[\deg\right]}\) & \(97.4_{-8.0}^{+12.1}\)& \(59.0_{-115.8}^{+55.5}\)& 97.0 & 3.5 & \derived & \derived \\
\(M_{A} \, \si{\left[\deg\right]}\) & \(347.3_{-14.7}^{+5.7}\) & \(134.0_{-55.6}^{+109.3}\) & 348.8 & 198.7 & \derived & \derived \\
\(\Delta \incl \, \si{\left[\deg\right]}\) & \dots & \(12.4_{-2.8}^{+3.5}\) & \dots & \(12.3\) & \dots  & \derived \\
\(\rho \, \left[\si[per-mode=power]{\gram\per\cm\cubed}\right]\) & 0.926$_{-0.136}^{+0.173}$ & \dots & 0.928 & \dots & \derived &  \dots \\
\(m_{p}~{\left[\mathrm{M}_{\mathrm{Jup}}\right]}\) & \(0.694_{-0.040}^{+0.038}\) & \(0.184_{-0.026}^{+0.022}\) & \(0.694\) & \(0.188\) & \derived & \derived \\
\(a_{p} \, \si{[\astronomicalunit]}\) & \(0.117_{-0.0014}^{+0.0014}\) & \(0.194_{-0.0023}^{+0.0022}\) & \(0.117\) & \(0.194\) & \derived & \derived \\
\bottomrule
\end{tabularx}
\tablecomments{The orbital elements are in the Jacobi frame and are valid for epoch BJD = 2458330.0. The adopted priors are listed in the right-most columns, and their meanings are \(\mathcal{U}\) -- uniform and \(\mathcal{N}\) -- Gaussian (normal) priors. The derived planetary posterior parameters of \(a\) and \(m\) are calculated, considering the stellar parameter uncertainties.}
\end{table*}

\subsection{RV and stellar activity spectroscopic data analysis}
\label{feros_analysis}
\noindent
As an integral part of our \wine\ efforts to validate and characterize warm \tess\ transiting exoplanets, we obtain, derive, and analyse precision RVs and activity index data from the \feros\ instrument. 
\autoref{MLP_results} shows a \gls\ periodogram of the extracted \feros\ spectroscopic data of \toinum\ as labeled in the sub-panels. 
The horizontal lines in the sub-panels denote the adopted periodograms \textsc{fap} (False Alarm Probability) thresholds of 10\%, 1\%, and 0.1\%; we consider signals surpassing the latter as significant. 
Vertical lines indicate the orbital periods of the two exoplanets \toinumb\ and \toinumc discovered in this work.

\autoref{MLP_results} shows that the \rv\ time series exhibits a forest of peaks near the \(\sim \qty[number-unit-product=\text{-}]{14.26}{day}\) period of the transiting planet, which we consider a promising indication of a coherent Doppler signal.  
In addition, we detect a significant linear trend in the \rv\ data, which we investigate further. 
As seen in the second panel of \autoref{MLP_results}, after applying a linear trend model to the \rv s, the residuals reveal a marginally significant periodic signal at \qty{14.25(2)}{\,\days}, fully consistent with the orbital period of the transiting planet. The semi-amplitude of this signal in the \feros\ data is \qty{50.62(5.55)}{\meter\second^{-1}}, suggesting a Jovian-mass companion with a minimum mass of approximately \(0.7\,M_{\rm Jup}\).

The residuals from our final joint \ttv+\rv\ model, including a linear trend component (see \autoref{Sec4.4}) are shown in the last panel of \autoref{MLP_results}, and these data show no remaining significant periodicity. Furthermore, the stellar activity indicators \bis\ (Bisector Inverse Slope), \fwhm, \(\rm{H}_{\alpha}\), He\,I  Na\,II do not exhibit significant periodic signals, as shown in the remaining panels of \autoref{MLP_results}.

The conclusions from the \feros\ spectroscopic analysis can be summarised as follows:  
(i) We detect a linear trend in the \rv s, suggesting the presence of an additional long-period planetary or substellar companion, which cannot be constrained with the current temporal baseline of the data.  
(ii) The \feros\ \rv s are sufficient to confirm the planetary nature of the transiting object \toinumb\ and to measure its dynamical mass.  
(iii) No conclusive evidence for the orbital period or mass of the outer perturber \toinumc\ can be inferred from the \feros\ data alone.  
(iv) There is no strong indication that \toinum\ is an active star, although higher-precision data and additional time series would be required to draw firm conclusions.
In this context, we note that based on the estimated \(v \sin(i)\) given in \autoref{table:phys_param}, the stellar rotation period is estimated to be \qty{13.1 \pm 1.1}{\,\days}, which is close to the mean orbital period of the transiting exoplanet \toinumb\ (cf. \autoref{TTV_plot1}). 
In case of significant stellar activity induced by rotational modulation, this spin-orbit proximity could have an impact in the extracted \rv\ signal and therefore potentially bias the derived semi-amplitude and thus its mass.

\subsection{Joint \ttv\ and RV analysis}
\label{Sec4.4}
\noindent
The preliminary Doppler data vetting revealed that the \feros\ measurements exhibit a linear \rv\ trend. 
After subtracting this component using a linear trend model, the \rv\ residuals exhibit a significant periodic signal at \(\sim \qty{14.25(2)}{\,\days}\), and a semi-amplitude of \qty{50.7(5.5)}{\meter\s^{-1}} (see \autoref{MLP_results}). 
This signal is consistent with the period of the transiting planet, supporting its interpretation as a giant-mass companion.

Consequently, we performed a joint fit analysis of the extracted \ttv s and the \feros\ \rv\ dataset using the \texttt{Exo-Striker} tool. 
We employ a self-consistent \(N\)-body dynamical model to fit simultaneously the data with the same fitting parameters. 
In our modeling scheme these parameters were the \rv\ semi-amplitude \(K_{\rm b,c}\), which was converted instantly to the dynamical planetary masses \((m_{p})_{\rm b,c}\), the osculating planetary orbital period \(P_{\rm b, c}\), the \rv\ offset \rv\ off. 
\feros, the \rv\ jitter parameter \rv\ jitt. \feros, and linear trend parameter \rv\(_{\rm{lin. tr. FEROS}}\) of the \feros\ \rv\ data. 
The planetary eccentricities \(e_{\rm b, c}\), arguments of periastron \(\omega_{\rm b, c}\), and mean anomalies \(M_{\rm b, c}\) were derived using the parameterization \(h_{\rm b, c} = e_{\rm b, c} \sin\left(\omega_{\rm b, c}\right)\), \(k_{\rm b, c} = e_{\rm b, c} \cos\left(\omega_{\rm b, c}\right)\), and \(\lambda_{\rm b, c} = \omega_{\rm b, c} + M_{\rm b, c}\), which we chose in \cite{Vitkova2025} since it is more efficient for orbits with small eccentricities \citep{Tan2013}.

Since we know that the perturber planet is not transiting, we further assumed a non-coplanar, mutually inclined orbital geometry and fitted for the orbital inclinations \(i_{\rm b,c}\) and the difference between the longitudes of the ascending nodes \(\Delta\Omega \equiv \Delta\Omega_{c - b} = \Omega_{c} - \Omega_{b}\), where the mutual inclination comes following the expression:
\begin{equation}
\Delta \incl = \arccos \! \big(\!\cos(i_c)\cos(i_{b}) + \sin(i_c)\sin(i_{b})\cos(\Delta\Omega)\big).
\label{eq:deltai}
\end{equation}

Our approach follows a similar \ttv+\rv\ N-body fitting methodology as used by \citet{Trifonov2021} for the \textsc{toi}--2202 system and,  more recently, for the \textsc{toi}--4504 system \citep{Vitkova2025}. 
Both \textsc{toi}--2202 and \textsc{toi}--4504 systems are very similar to \toinum, since all these systems consist of a transiting warm Jovian with a strong \ttv\ signal, and a non-transiting perturber exoplanet found close to the 2:1 \mmr\ commensurability with the transiting planet. 
We refer the reader to \citet{Trifonov2021} and \citet{Vitkova2025} for further details.

A global orbital fit was performed using broad priors and the \texttt{dynesty} sampler. 
We employed 100 ``live points" per fitted parameter, focusing on the convergence of the Bayesian log-evidence, and utilized a ``static" nested sampler \citep[see][for details]{Speagle2020}. 
For \toinumb, prior parameter ranges were adopted following the results of our \ttv\ extraction models and the \gls\ analysis of the \feros\ \rv s. 
For the perturber \toinumc, we explored a broad parameter space encompassing eccentricities, masses, and orbital periods. 
Due to the large and ambiguous parameter space that must be explored, particularly given that transit timing variations (\ttv s) can often lead to highly degenerate solutions \citep[e.g.,][]{Lithwick2012}, and the additional complexity introduced by fitting a non-transiting planet \citep[e.g.,][]{Yahalomi2024}, we initially performed a broad parameter scan using a nested sampling (\ns) scheme with wide, non-informative (flat) priors. 

Specifically, we search for interior non-transiting planets with periods in the range \(P_{\rm c} \in \mathcal{U}(4.0, 10.0)\) days, \(e_{\rm c} \in \mathcal{U} (0.0, 0.2)\), and \(m_{\rm c} \in \mathcal{U} (0.01, 0.5) \, \Mjup\). 
These prior ranges did not lead to adequate solutions which could explain the \rv s and \ttv s. 
Further, we test exterior planets, with periods \(P_{\rm c} \in \mathcal{U}(20.0, 60.0)\) days, which have provided promising fit solutions that we examine in more depth. 
The exact wide prior ranges for all parameters in this preliminary parameter scan were defined experimentally in an iterative fashion until we set well-defined priors for an efficient \ns\ test, and shall not be discussed here.

After inspecting the resulting (multimodal) parameter posteriors, we identified the samples with the highest posterior probability. 
We then performed a Simplex likelihood optimization, starting from the nested sampling (\ns) point with the highest \(-\ln\mathcal{L}\) value to refine the best-fit solution. 
Finally, we conducted an \mcmc\ analysis around the optimized Simplex solution to construct posterior probability distributions and derive the \(1\sigma\), \(2\sigma\), and \(3\sigma\) confidence intervals for the model parameters consistent with the available data.

\begin{figure*}[t]
\includegraphics[scale=1]{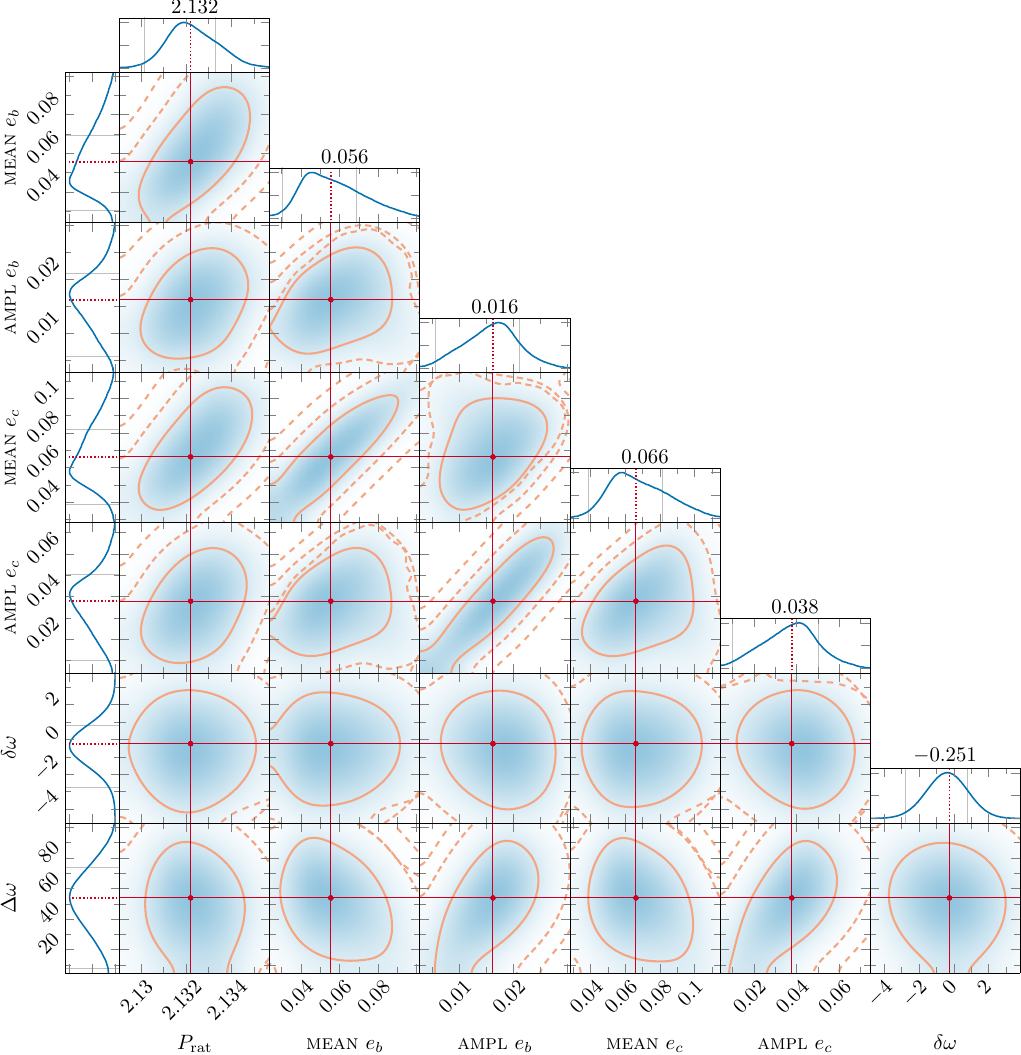} 
\caption{Posterior distributions of the key dynamical parameters near the 2:1 mean-motion resonance (\mmr) for the two-planet system, derived from a combined \ttv+\rv\ \(N\)-body model. A total of 10,000 samples were drawn from the posterior distribution and individually tested for long-term stability. The distributions reflect the system's behavior in the vicinity of the 2:1 period ratio, highlighting the resonant and secular dynamics. We report the median values and peak-to-peak variations of the period ratio (\(P_{\rm rat}\)), eccentricities (\(e_{\rm b}\) and \(e_{\rm c}\)), and apsidal angle difference (\(\Delta\omega\)), along with their dynamical amplitudes: amplitude of \(e_{\rm b}\) and of \(e_{\rm c}\), and \(\delta\omega\). Dynamically constrained planetary masses and semi-major axes are also shown. The resonant angles \(\theta_{1}\) and \(\theta_{2}\) are omitted as their evolution is dominated by circulation over the full \qtyrange{0}{360}{\degree} range. Contours in the 2D distributions indicate the \(1\sigma\), \(2\sigma\), and \(3\sigma\) confidence intervals of the stable posterior set.
}
\label{dyn_samp} 
\end{figure*}

The best dynamical fit to the \ttv\ and \rv\ data is shown in \autoref{TTV_plot2}, whereas our final results are summarized in \autoref{table:NS_params}, listing our final \mcmc\ priors, posteriors, and best \(-\ln\mathcal{L}\) solution consistent with the \ttv\ and \rv\ data. 
Additionally, the \mcmc\ posterior plot is shown in \autoref{mcmc_samp}. 
Analysis of the \mcmc\ posterior probability distribution suggests that the \ttv s of \toinumb\ are most likely caused by an exterior Saturn-mass planet with an orbital period near the 2:1 ratio. 
Our analysis of \ttv s, combined with \rv\ data from \feros, robustly indicates the presence of a compact massive Jovian-Saturn mass pair of planets. 
The system consists of two well-characterized companions with orbital periods of \(P_{\rm b} = 14.256_{-0.001}^{+0.001}\) days and \(P_{\rm c} = 30.356_{-0.012}^{+0.010}\) days. 
Their eccentricities are tightly constrained at \(e_{\rm b} = 0.065_{-0.013} ^{+0.020}\) and \(e_{\rm c} = 0.027_{-0.016}^{+0.027}\), while their dynamical masses are determined to be \(m_{\rm b} = 0.694_{-0.040}^{+0.038} \, \Mjup\) and \(m_{\rm c} = 0.194_{-0.023}^{+0.022} \, \Mjup\). 
Despite being close to the 2:1 \mmr\ commensurability, we will show in \autoref{sec5} that the system does not reside in a 2:1 \mmr. 
We obtained an average period ratio of \(\sim 2.13\), which is too far for a liberating eccentricity-type resonance, given the estimated small eccentricities. 
A similar dynamical architecture has been observed in the \textsc{toi}--2202\,b \& c system \citep{Trifonov2021}, \textsc{toi}--2525\,b \& c system \citep{Trifonov2023}, which are close but not in the 2:1 \mmr.

\begin{figure*}[!ht]
    \includegraphics[scale=1]{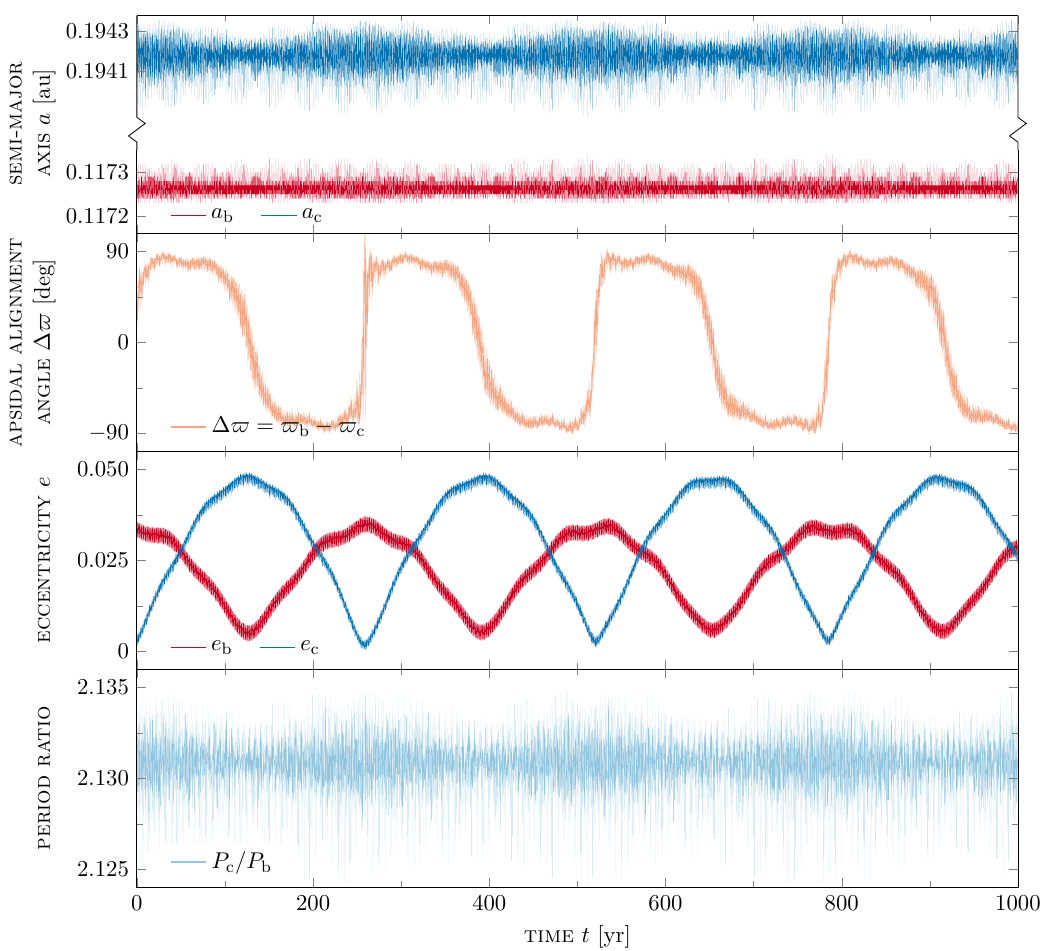} 
    \caption{Orbital evolution of the \toinum\ system for a 1000--year--long \(N\)-body integration using Wisdom-Holman scheme. \emph{From top to bottom:}
    \emph{Panel 1.} Evolution of the planetary semi-major axes \(a_{\rm{b}}\) and \(a_{\rm{c}}\). 
    \emph{Panel 2.} Evolution of the apsidal alignment angle \(\Delta\varpi = \varpi_{\rm b} - \varpi_{\rm c}\). 
    \emph{Panel 3.} Evolution of the planetary eccentricities \(e_{\rm{b}}\) and \(e_{\rm{c}}\). 
    \emph{Panel 4.} Evolution of the period ratio \(P_{\rm{c}} / P_{\rm{b}}\). 
    \emph{Panel 5.} Evolution of the angle \(\theta_{1}\). 
    The system \toinum\ is outside of the 2:1 \mmr, as no libration in any of the resonance angles is observed and the mean period evolution is osculating significantly above the 2:1 period ratio.\label{evol_plot}}
\end{figure*}

We also note that our \ttv+\rv\ joint model includes a linear \rv\ trend component, which hints at the presence of an additional massive and long-period companion orbiting \toinum. 
Although the origin of this sparsely sampled, yet statistically significant trend is beyond the scope of this study, we performed a crude estimate of the minimum masses and semi-major axes of potential companions that could produce the observed slope. 
Assuming nearly circular orbits, we find that sub-stellar objects with masses ranging from a few Jupiter masses up to low-mass M dwarfs, and semi-major axes between 12 and \qty{40}{\astronomicalunit}, could be responsible. 
While such a distant companion would not significantly affect the observed \ttv s or \rv s in a way that bias our conclusions, it could have important implications for the formation, evolution, and long-term dynamical architecture of the inner planetary system.

\begin{figure*}[t]
\includegraphics[scale=1]{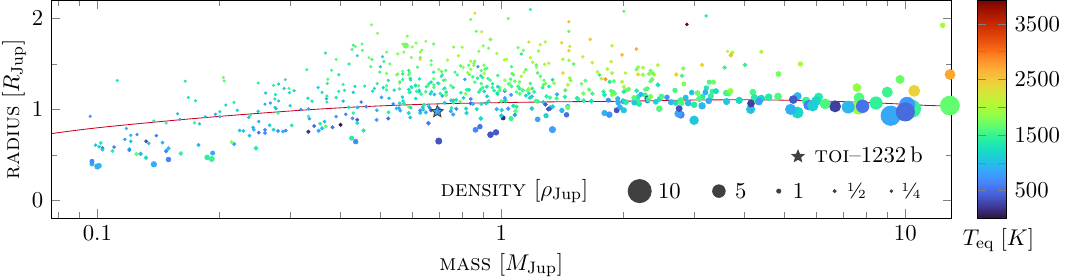}
\caption{Radius–mass distribution of all giant transiting planets with measured masses from the TepCat catalog. The masses were accessed either by \ttv s, \rv s, or both. The position of \toinumb\ is marked with a star symbol.
Color-coded is the equilibrium temperature. The solid curve indicates the predicted planetary radius based on the models of \citet{Fortney2007} and consistent with the stellar parameters of \toinum.}
\label{fig:rho} 
\end{figure*}

\section{Dynamics and long-term stability}
\label{sec5}
\noindent
In this section we investigate the dynamical configuration implied by our joint \ttv+\rv\ solution and assess whether the system can remain stable over long timescales. 
Using posterior samples from the global fit, we perform direct \(N\)-body integrations to test long-term stability and to diagnose the resonant and secular behavior of the \toinum\ planet pair near the 2:1 resonance. 
We then use the dynamically consistent posterior set to provide forward transit-time predictions for future follow-up observations.

\subsection{Numerical simulations}
\label{sec5.1}
\noindent
We conducted a comprehensive \(N\)-body dynamical analysis of the \toinum\ system using the symplectic \(N\)-body integrator \symba\ \citep{Duncan1998}.
The \symba\ integrator is designed to accurately handle close planetary encounters during simulations, while maintaining accuracy and efficiency. 
Our implementation of \symba\ is directly integrated into the \texttt{Exo-Striker} software and has been customized to operate in Jacobi elements \citep[e.g.,][]{Lee2003}, making it compatible with direct input from best-fit parameters and posterior samples.

The tightly packed orbital configuration of the \toinum\ planets necessitated the use of a relatively small initial integration time step of 0.2 days to accurately compute their orbital evolution. 
We carried out long-term dynamical stability simulations of the \toinum\ system over \qty{10}{\mega\year} for a set of 1,000 randomly selected samples drawn from the \mcmc\ posterior distributions. 
For each simulation, we monitored the time evolution of the planetary semi-major axes and eccentricities to verify that the system remained dynamically stable and well-separated. 
We classified a system as dynamically unstable if any planet’s semi-major axis deviated by more than 20\% from its initial value, or if eccentricity growth led to orbit-crossing configurations. 
Similar simulation setups using \symba\ and the \texttt{Exo-Striker} have been applied to other \ttv--\rv\ systems discovered by \tess\ and analyzed by our \wine\ team \citep[e.g., see][]{Trifonov2021, Trifonov2023, Hobson2023, Bozhilov2023, Vitkova2025}.

Given the period ratio of the system close to the 2:1 commensurability, as a rule in our dynamical analysis scheme, we inspect the first-order \mmr\ angles \(\theta_{1}\) and \(\theta_{2}\), which are defined as:
\begin{equation}
\theta_{1} = \lambda_{\rm b} - 2 \lambda_{\rm c} + \varpi_{\rm b}, \quad
\theta_{2} = \lambda_{\rm b} - 2 \lambda_{\rm c} + \varpi_{\rm c},
\end{equation}
where \(\varpi_{\rm b, c} = \Omega_{\rm b, c} + \omega_{\rm b,c}\) are the planetary longitudes of periastron and \(\lambda_{\rm b, c} = {M_{0}}_{\rm b,c} + \varpi_{\rm b, c}\) are the mean longitudes, respectively. 
We also monitored the evolution of the secular apsidal angle \(\Delta\omega\), which is defined as:
\begin{equation}
\Delta\omega = \theta_{1} - \theta_{2} = \varpi_{\rm b} - \varpi_{\rm c},
\end{equation}
which indicates whether secular interactions dominate the dynamics of the system. 

The results from our long-term stability analysis indicate that all examined 1,000 samples are stable for \qty{10}{\mega\year} with very similar dynamical behavior. 
\autoref{dyn_samp} shows the derived posteriors of the dynamical properties of the studied 1,000 samples. 
The distribution of dynamical parameters reveals low-eccentricity evolution, but despite being close to the 2:1 \mmr, the \toinum\ pair seems to reside outside of the low-order eccentricity type 2:1 \mmr. 
The mean period ratio evolution is oscillating around 2.13, while we did not detect libration of the resonance angles \(\theta_{1}\), \(\theta_{2}\), the apsidal alignment angle \(\Delta\omega\) show libration around \qty{0}{\degree} for about 80\% of the samples. 
The posterior distributions of \(\theta_{1}\), \(\theta_{2}\) are consistent with circulation, libration amplitudes between \qty{0}{\degree} and \qty{360}{\degree}, whereas \(\Delta\omega\) is consistent with large semi-amplitudes of about \qty{60}{\degree}.

\autoref{evol_plot} presents a 1000\,yr snapshot of the dynamical evolution for the best fit corresponding to the maximum \(-\ln\mathcal{L}\) value from the posterior probability samples. 
\autoref{evol_plot} from left to right, illustrates the evolution of the semi-major axes \(a_{\rm b}\) and \(a_{\rm c}\), the orbital eccentricities \(e_{\rm b}\) and \(e_{\rm c}\), and evolution of the period ratio, respectively. 
No libration of the eccentricity-type first-order resonance angles was detected, nor is there any libration around a fixed point in the trajectory evolution that would indicate the presence of a 2:1 \mmr. 
 
\subsection{Follow-up transit predictions}
\noindent
\toinum\ is a good target for a follow-up transit photometric monitoring from the Southern hemisphere. 
Using our posterior analysis, we predicted the upcoming transit the mean (osculating) period of \toinumb\ of \(P_{\rm b} = 14.25503_{-0.0007}^{+0.0008}\) days.
We note that the overall precision of the \ttv\ predictions deteriorates over time; therefore, we provide transit predictions only up to 2030. 
We provide our transit prediction estimates in \autoref{tab:transit_times}.

\section{Internal composition of \toinumb}
\label{sec6}
\noindent
We examine the standing of the transiting exoplanet \toinumb\ within the broader population of exoplanets for which masses, radii, and density have been accessed reliably via transits, \rv\ measurements, and \ttv s. 
In \autoref{fig:rho}, we show the exoplanet mass–radius distribution limited to Jovian giants with measured masses from 0.05 to \qty{11}{\MJup} and radius between 0.15 and \qty{3.0}{\RJup}. 
Additional dimensions in \autoref{fig:rho} are the color-coded estimated planetary equilibrium temperature, and symbol-size coded by the planetary mean density. 
We include a mass–radius model from \citet{Fortney2007}, interpolated using the estimated stellar luminosity of \toinum, a semi-major axis of \qty{0.117}{\astronomicalunit}, and an assumed system age of \qty{4.0}{\giga\year}. 
With a mean density posterior estimate of \(\rho_{\rm b} = 0.926_{-0.136}^{+0.173} \; \si{\gram\per\cm\cubed}\), \toinumb\ is consistent with the bulk of giant planets with measured radius.

\begin{figure}[b]
\includegraphics[scale=1]{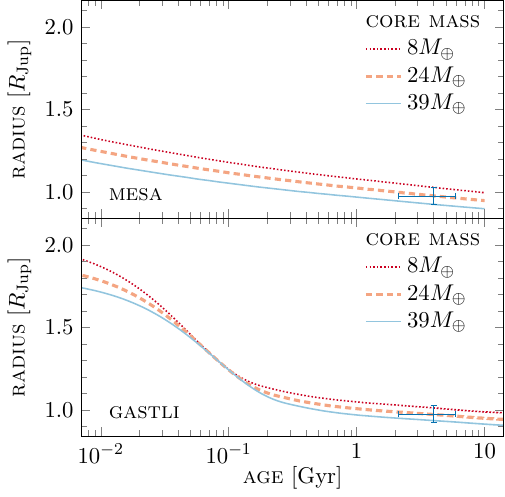}
\caption{Age-radius diagram using \mesa\ (\emph{Upper panel}) and \gastli\ (\emph{Lower panel}) models with the position of \toinumb\ shown in black. Over-plotted are three different models with a fixed envelope metallicity (\(Z = 0.023\)), and varying core masses.}
\label{fig:age_radius} 
\end{figure}

We provide internal composition estimates for the transiting planet \toinumb. 
For this purpose, we computed planet interior models and their thermal evolution using the Modules for Experiments in Stellar Astrophysics (\mesa) \citep{paxton2011, paxton2013}, following the methods described in \cite{Jones2024}. 
In this case, we modeled the planet with an inert isodensity core with different masses, surrounded by a gaseous envelope with the same host star metallicity (\(Z = 0.023\)). 
For the core, we assumed a 1:1 mixture of rock and ice, with their density obtained from the \(\rho-P\) relations presented in \cite{Hubbard1989}. 
We evolved different models, with different masses of the core, and we compared them with the current position of \toinumb\ in the age--radius diagram. 
The top panel of \autoref{fig:age_radius} shows different models that agree at the \(1-\sigma\) level with the current radius of the planet. 
These results correspond to a planetary metallicity of \(Z_{p} = 0.13_{-0.7}^{+0.7}\), and a corresponding heavy-element enrichment with respect to the host star of \(Z_{p} / Z_{\star} = 9.6_{-4.1}^{+3.1}\).

For comparison, we also computed interior models for \toinumb\ using the \textsc{gas} gian\textsc{t} mode\textsc{l} for \textsc{i}nteriors \citep[\gastli,][]{acuna2021, acuna2024}, whose results are shown in bottom panel of \autoref{fig:age_radius}. 
\gastli\ assumes a core composed of a 1:1 water-silicate mixture and an envelope of water, helium, and hydrogen. 
The metallicity of the envelope is a free parameter that can range from subsolar to 250 x solar with the \gastli\ default atmospheric grid. 
For our models, we adopt a solar envelope. 
In contrast to \mesa, \gastli\ calculates the density and temperature of the core self-consistently given the equations of state of rock, water and \(\text{\emph{H}}/\text{\emph{He}}\) ratio. 
This difference is negligible for sub-Neptunes, but can have a significant effect for planets exceeding a pressure of \(\SI{e10}{\pascal}\) in the core-envelope boundary \citep{Mordasini12, ChenRogers2016}. 
The global average equilibrium temperature depends on the stellar parameters and is estimated to be \(T_{\rm{eq}} \approx \qty{849}{\kelvin}\), assuming null albedo:
\begin{equation}
    T_{\mathrm{eq}}^{4} = \frac{T_{\star}^{4}}{4} \left(\! \frac{R_{\star}}{a_{\mathrm{d}}} \!\right)^{\!2},
\end{equation}
where \(T_{\star}\) and \(R_{\star}\) are the star's effective temperature and radius and \(a_{\mathrm{d}}\) is the orbital semi-major axis.

We computed the age--radius relationship for five different core mass assumptions, ranging from 5 to \qty{102}{\MEarth}, and compared the results with the actual position of \toinumb\ in the diagram. 
The results are presented in \autoref{fig:age_radius} and agree well with the \mesa\ output for the age of \toinumb. 
The models differ for ages \(\lesssim \qty{100}{\mega\year}\) because they use different values for the initial entropy: \(S_{0} = 12 \, k_{B} m_{H}\) for \gastli\ models and \(S_{0} = 9.6 \, k_{B} m_{H}\) for \mesa\ models.\footnote{Here \(k_{B} = \SI[scientific-notation=true]{1.3806e-23}{\joule\per\kelvin}\) is the Boltzmann constant and \(m_{H} = \SI[scientific-notation=true]{1.6735e-27}{\kilo\gram}\) is the mass of the Hydrogen atom.}

\section{Summary and Conclusions}
\label{sec7}
\noindent
In this work, we report a new multiple exoplanet system around \toinum. 
This discovery was made as part of the \wine\ survey, which efficiently validates \tess\ transiting warm Jovian planets around relatively bright stars with \rv\ follow-up. 
Further, \wine\ is very effective in recovering the orbital and physical parameters of multiple planet systems which show significant \ttv\ signals.

The \toinum\ system is such an example. 
We detected a compact Jovian-mass pair of planets around the G-dwarf star \toinum, which we estimate to have a stellar mass of \(1.058_{-0.063}^{+0.065} \, M_{\odot}\) and a radius of \(1.072_{-0.047}^{+0.047} \, R_{\odot}\). 
\tess\ revealed the transiting warm Jovian mass planet \toinumb\, which transits with a period of \(14.256_{-0.001}^{+0.001}\) days. 
\tess\ transits together with ground-based follow-up light curves clearly show significant \ttv s with a semi-amplitude of about \qty{27}{\min}, suggesting the presence of a second, non-transiting, massive body which perturbs the transiting planet.

Further, we note that given the significant linear trend in the \rv s, additional non-transiting planets could reside in the \toinum\ system.
 
Our dynamical analysis of the system configuration revealed that the new Gas-giant pair osculates outside the exact 2:1 \mmr\, librating configuration. 
Based on the available transit and \rv\ data we could rule out a 2:1 \mmr\ since the orbital configuration we obtain and the forward \(N\)-body simulations showed that the osculating period ratio of \toinum\ is strictly larger than two, similarly to other \tess--\ttv\ systems such as \textsc{toi}--2202, and \textsc{toi}--2525, but is also consistent with the peak of the distribution of period ratios of planet pairs just above 2 observed by Kepler \citep{Lissauer2011, Fabrycky2014}. Most Kepler planets are significantly smaller and less massive than \toinumbc, resulting in a much narrower dynamical width of the 2:1 \mmr. Consequently, the Kepler peak may also lie slightly beyond the resonance libration zone. It is worth noting that only a small fraction of Kepler systems have sufficiently constrained dynamical solutions to exclude resonance libration based on eccentricity estimates.

The \toinum\ system is interesting because of the warm Jovian and Saturn-mass planet pair near the 2:1 \mmr, which is a rarely observed configuration, and their formation and dynamical evolution are still not well understood. 
Thus, the \toinum\ system is yet another important addition to the strong \tess-\ttv\ systems, and it will help to better understand the planetary formation and evolution.

\begin{acknowledgments}
\noindent
This research has made use of the Exoplanet Follow-up Observation Program website, which is operated by the California Institute of Technology, under contract with the National Aeronautics and Space Administration under the Exoplanet Exploration Program. 
We acknowledge the use of public \tess\ data from pipelines at the \tess\ Science Office and at the \tess\ Science Processing Operations Center. 
Funding for the \tess\ mission is provided by \nasa's Science Mission directorate.

% MAST
This paper made use of data collected by the \tess\ mission and are publicly available from the Mikulski Archive for Space Telescopes (\textsc{mast}) operated by the Space Telescope Science Institute (\textsc{sts}c\textsc{i}).
Some of the data presented in this paper were obtained from the Mikulski Archive.

% ASTEP
The \astep\ team thanks the dedication and technical support of the entire French Polar Agency (\textsc{ipev}). 
We also wish to thank the technical staff at Concordia Station, and give a particular recognition to the work and efforts produced by the entire wintering crew at Concordia to ensure a continuity of operations throughout each Antarctic winter. 
\astep\ benefited from the support of the French and Italian polar agencies \textsc{ipev} and \textsc{pnra} in the framework of the Concordia station program and from \textsc{oca}, \textsc{insu}, Idex \textsc{ucajedi} (\textsc{anr}-15-\textsc{idex}-01) and \textsc{esa} through the 
\textsc{s}cience \textsc{f}aculty of the \textsc{e}uropean \textsc{s}pace \textsc{r}esearch and \textsc{t}echnology \textsc{c}entre (\textsc{estec}). 
This research received funding from the \textsc{e}uropean \textsc{r}esearch \textsc{c}ouncil (\textsc{erc}) under the \textsc{e}uropean \textsc{u}nion's \textsc{h}orizon 2020 research and innovation programme (grant agreement No. 803193/\textsc{bebop}), and from the \textsc{s}cience and \textsc{t}echnology \textsc{f}acilities \textsc{c}ouncil (\textsc{stfc}, grant No. \textsc{st}/\textsc{s}00193\textsc{x}/1, \textsc{st}/\textsc{w}002582/1, and \textsc{st}/\textsc{y}001710/1).

% SPOC
Resources supporting this work were provided by the \nasa\ High-End Computing (\textsc{hec}) Program through the \nasa\ Advanced Supercomputing (\textsc{nas}) Division at Ames Research Center for the production of the \spoc\ data products.

This work makes use of observations from the Las Cumbres Observatory global telescope network. 
The authors acknowledge support from the Swiss NCCR PlanetS and the Swiss National Science Foundation. 
This work has been carried out within the framework of the NCCR PlanetS supported by the Swiss National Science Foundation under grants 51NF40182901 and 51NF40205606.

D. P. M. acknowledges support by the \bnsf\ program ``\vihren--2024" project No. KP--06--DV/9/17.12.2024 and also support from the 2024 MPIA summer visitors program. 
T. T., S. S, V. B., and D. S. acknowledge support by the Bulgarian National Science Fund (\bnsf) program ``\vihren--2021" project No. KP--06--DV/5/15.12.2021. 
R. B. acknowledges support from \textsc{fondecyt} Project 1241963 and from \textsc{anid} -- Millennium  Science  Initiative -- ICN12\_009.
A. J. acknowledged support from \textsc{anid} -- Millennium  Science  Initiative ICN12\_009, IM23-0001 and \textsc{fondecyt} project 1251439.
H. P. acknowledges support by the Spanish Ministry of Science and Innovation with the Ramon y Cajal fellowship number RYC2021-031798-I, and funding from the University of La Laguna and the Spanish Ministry of Universities.
J. K. acknowledges support from the
Swedish National Space Agency (SNSA; DNR 2020-00104) and of the Swiss
National Science Foundation under grant number TMSGI2\_211697.

This work has been carried out within the framework of the National Centre of Competence in Research PlanetS supported by the Swiss National Science Foundation.
\end{acknowledgments}

\software{
    \texttt{Exo-Striker} \citep{Trifonov2019_es},
    \texttt{ceres} \citep{ceres},
    \texttt{zaspe} \citep{zaspe},
    \tesseract\ (Rojas, in prep.),
    \texttt{TESSCut} \citep{TESSCut},
    \lightkurve \citep{lightkurve},
    \texttt{corner} \citep{corner},
	\texttt{dynesty} \citep{Speagle2020},
    \texttt{batman} \citep{Kreidberg2015},
    \texttt{celerite} \citep{celerite},
    \texttt{wotan} \citep{Hippke2019},
    \texttt{transitleastsquares}  \citep{Hippke2019b},
}

\bibliographystyle{aasjournal}
\bibliography{main}

\appendix
\setcounter{table}{0}
\renewcommand{\thetable}{A\arabic{table}}

\setcounter{figure}{0}
\renewcommand{\thefigure}{A\arabic{figure}}

\begin{figure*}[tp]
\centering
\includegraphics[scale=1]{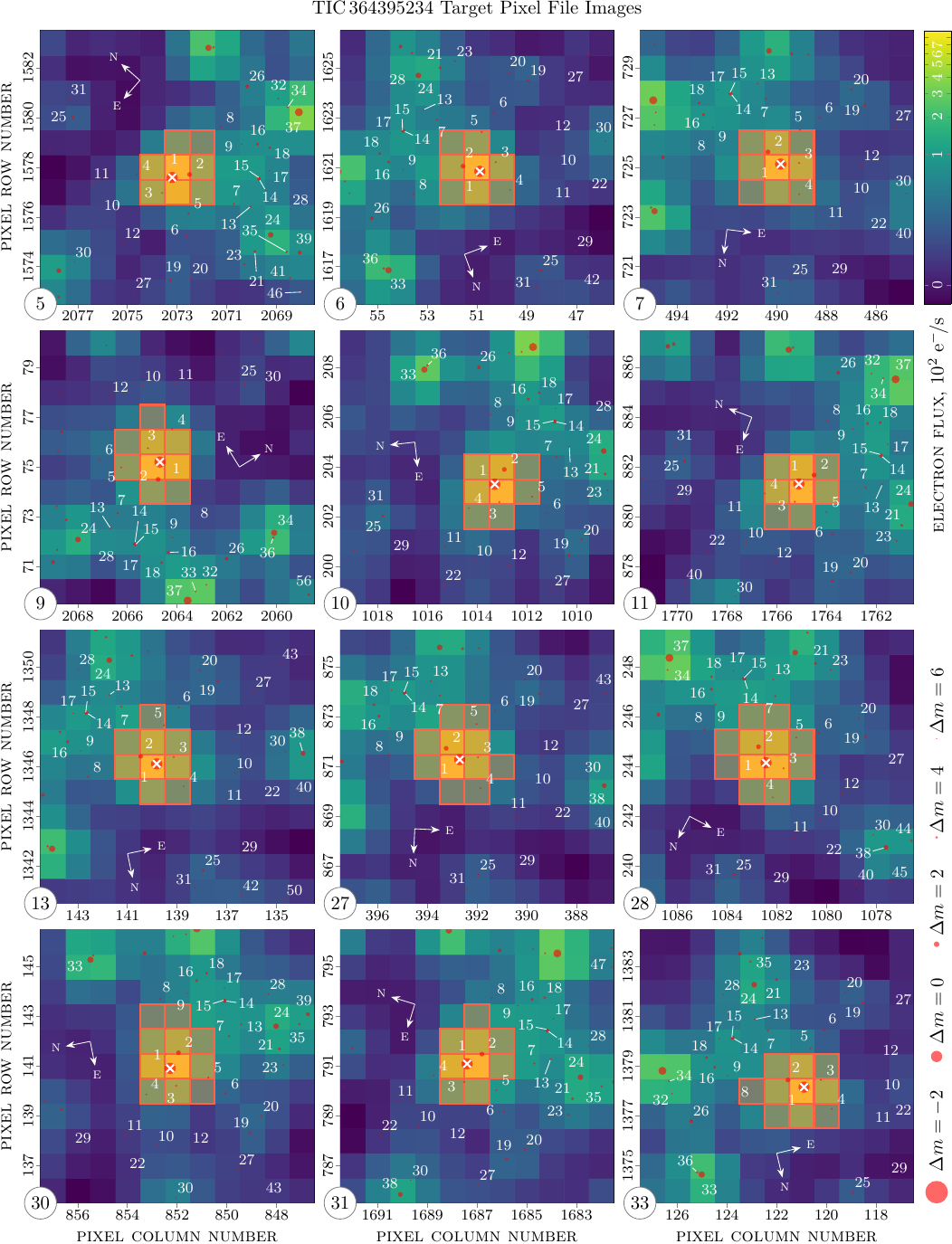}
    \caption{Target pixel file (\tpf) image of \toinum\ in \tess\ Sectors 5, 6, 7, 9, 10, 11, 13, 27, 28, 30, 31, and 33. The red dots represent the position of the \gaia\ sources in the field. \toinum\ resides in the middle, marked with a white \(\times\). The pixels marked with red borders are the ones used to construct the \tess\ Simple Aperture Photometry (\sap).}
\label{fig:tpf2}
\end{figure*}

\begin{figure*}[tp]
\centering
\includegraphics[scale=1]{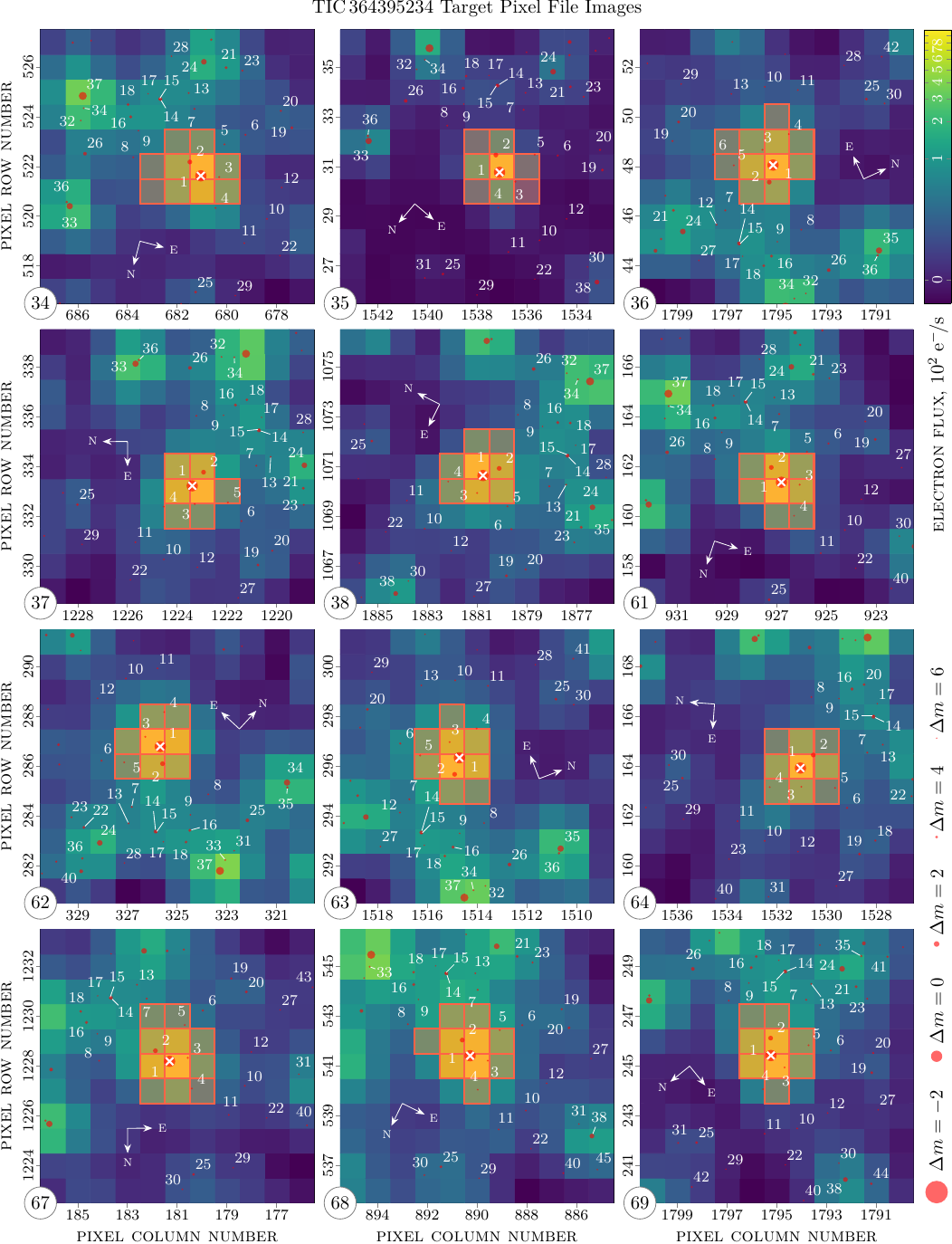}
    \caption{Target pixel file (\tpf) image of \toinum\ in \tess\ Sectors 34, 35, 36, 37, 38, 61, 62, 63, 64, 67, 68, and 69. The red dots represent the position of the \gaia\ sources in the field. \toinum\ resides in the middle, marked with a white \(\times\). The pixels marked with red borders are the ones used to construct the \tess\ Simple Aperture Photometry (\sap).}
\label{fig:tpf3} 
\end{figure*}

\noindent
Figures \ref{fig:tpf2} and \ref{fig:tpf3} show the full set of \tess\ target pixel file (\tpf) images of \toinum\ for all additional sectors. 
The figures display the positions of nearby \gaia\ \textsc{dr}3 sources relative to the target star and the \tess\ Simple Aperture Photometry (\sap) masks used in the extraction of the light curves. 
These images visually demonstrate the crowded photometric environment and the presence of nearby sources capable of contaminating the \tess\ aperture.

\autoref{fig:relfluxes} shows the undetrended \tess\ light curves of \toinum\ across all observed sectors. 
The transit events of \toinumb\ are marked in red, while the eclipses of the contaminating eclipsing binary are marked in blue.

\begin{figure*}[tp]
    \includegraphics[scale=1]{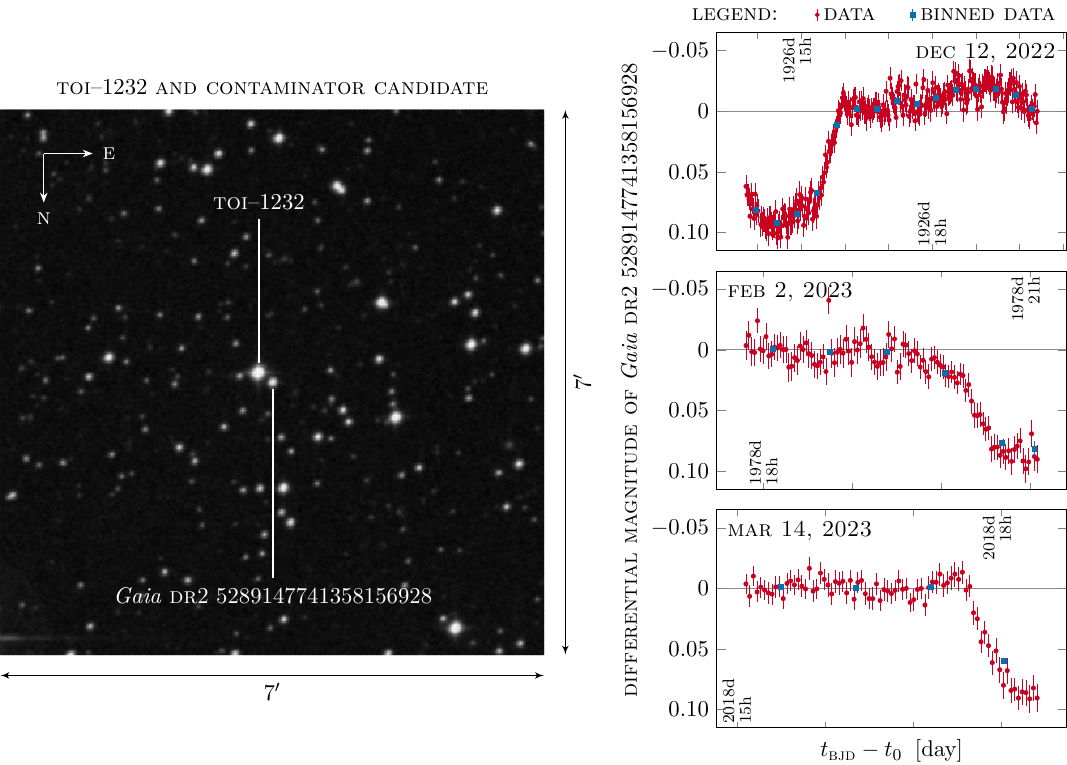}
\caption{Light curves of the contaminator candidate \gaianum, observed by the 60cm telescope at El Sauce, showing binary signal eclipses at three predicted times.}
\label{fig_contaminator} 
\end{figure*}

\autoref{fig_contaminator} shows ground-based follow-up photometry of the contaminating source \gaianum, obtained with the 60 cm telescope at El Sauce Observatory. 
The deep eclipses observed at the predicted times conclusively confirm the eclipsing binary nature of this nearby star and its role as the source of the short-period signal detected in the \tess\ data.

\begin{figure*}[tp]
\includegraphics[scale=1]{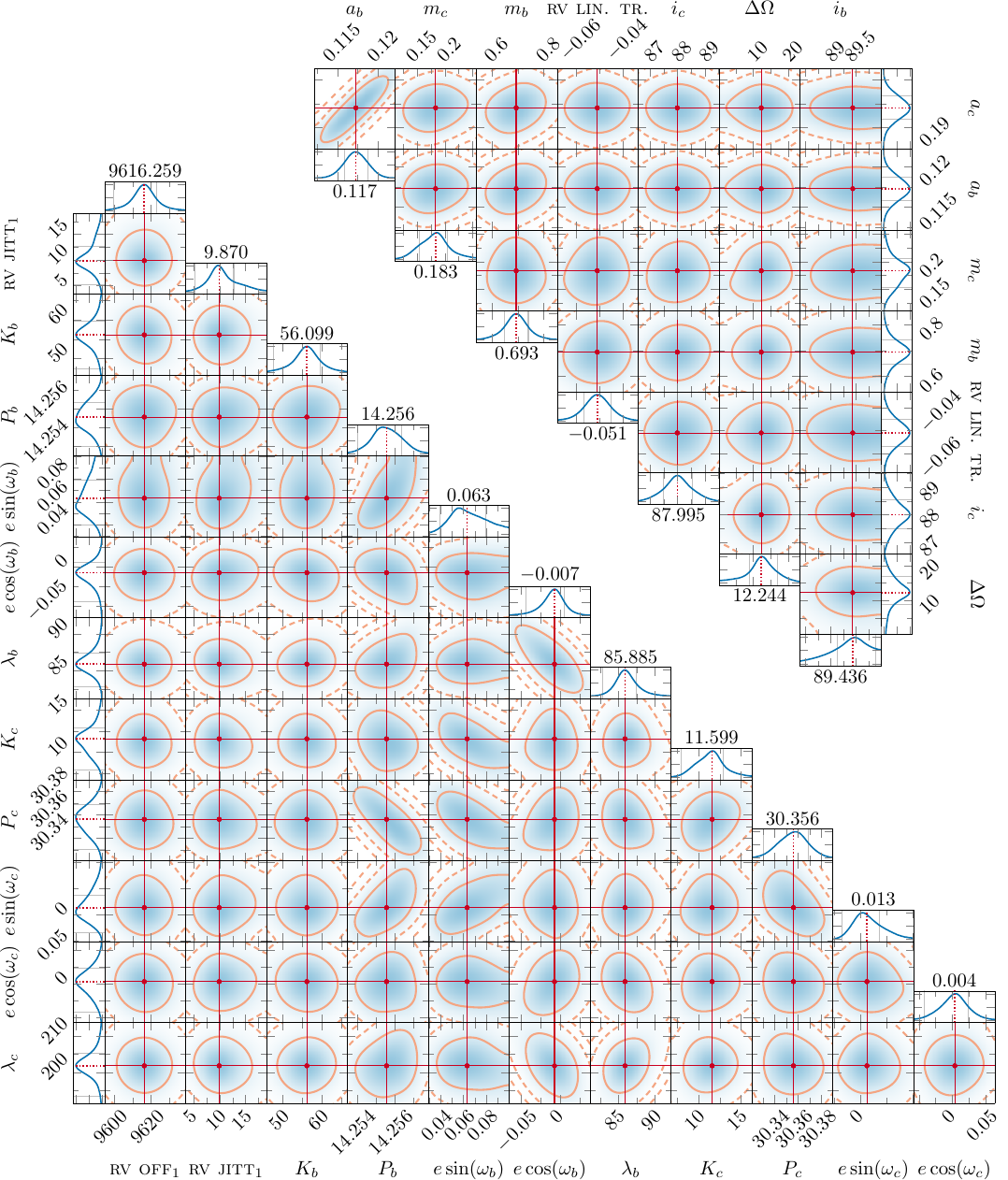} 
\caption{\mcmc\ posterior distribution from a \ttv+\rv\ fit. 
The median values of the fitted parameters and those derived are marked in red. The contours in each panel denote the \(1\sigma\), \(2\sigma\), and \(3\sigma\) regions of the distribution.}
\label{mcmc_samp} 
\end{figure*}

\autoref{mcmc_samp} displays the full \mcmc\ posterior distributions resulting from the joint \ttv+\rv\ modeling. 
The median values of the fitted and derived parameters are indicated for reference.

Finally, we include complete tables of extracted mid-transit times, transit model parameters, stellar activity indicators, and radial velocity measurements.

\vspace*{100pt}

\begin{figure*}[t!]
    \centering
    \begin{subfigure}{\textwidth}
        \includegraphics[scale=1]{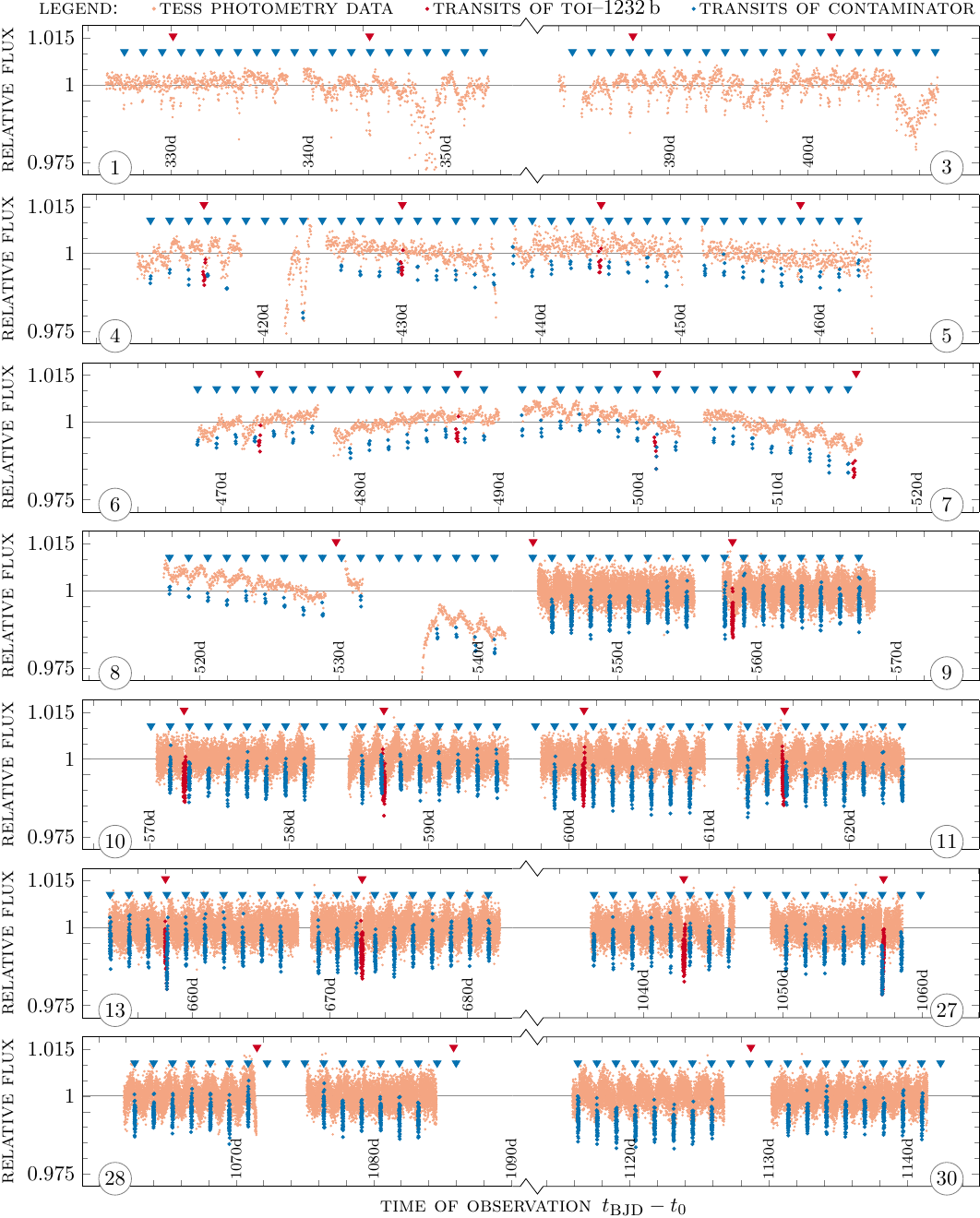}
        \subcaption{Undetrended \tess\ photometry data of \toinum\ for sectors 1, 3, 4, 5, 6, 7, 9, 11, 13, 27, 28 and 30. Marked in red are the transits of \toinumb. The eclipsing events of the contaminator are marked in blue and have been identified with the \texttt{wotan} package for all sectors except sectors 1 and 3, where they were located manually based on extrapolations from sector 4.}
        \label{fig:relflux1}
    \end{subfigure}
    \caption{Undetrended \tess\ photometry data of \toinum.}
\end{figure*}%
\begin{figure*}[t!]\ContinuedFloat
    \centering
    \begin{subfigure}{\textwidth}
        \includegraphics[scale=1]{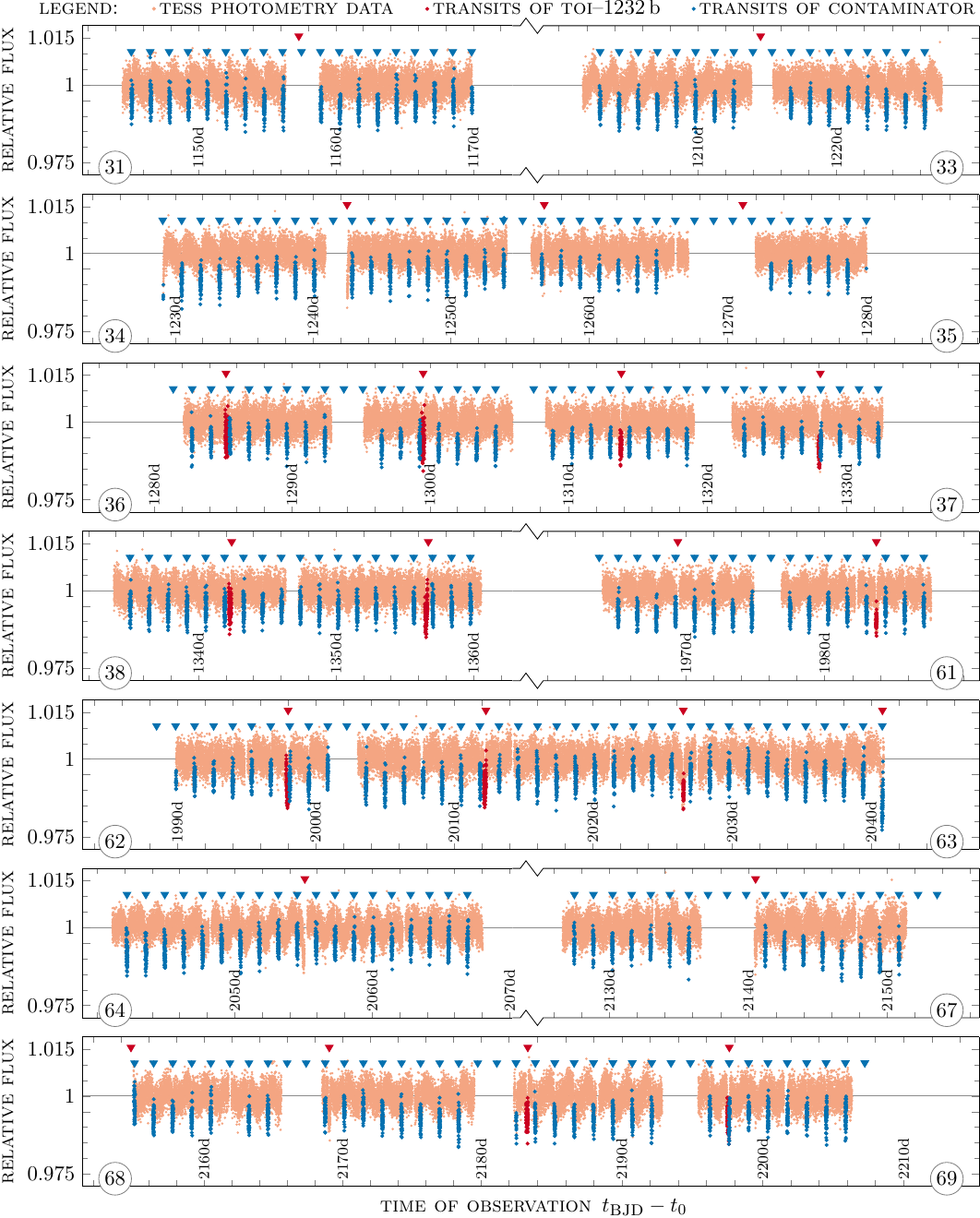}
        \subcaption{The plot shows the undetrended \tess\ photometry data of \toinum\ for sectors 31, 33, 34, 35, 36, 37, 38, 61, 62, 63, 64, 67, 68, and 69. Marked in red are the transits of \toinumb. The eclipsing events of the contaminator are marked in blue and have been identified with the \texttt{wotan} package for all sectors.}
        \label{fig:relflux2}
    \end{subfigure}
    \caption{Undetrended \tess\ photometry data of \toinum.}
    \label{fig:relfluxes}
\end{figure*}

\begin{table}[!hb]
    \caption{Mid-transit time estimates of \toinumb\ extracted from \tess\ and used for TTV analysis. The estimates are all provided relative to an initial time \(t_{0} = 2458000.0\) days.\label{table:TTVdataUpdated}}
\begin{tabularx}{\textwidth}{r r@{\hskip 19pt} r r@{\hskip 19pt} r r@{\hskip 19pt} r r@{\hskip 19pt} r r}
    \toprule
    \# & \(t_{\textsc{bjd}} - t_{0}\) [day] & \# & \(t_{\textsc{bjd}} - t_{0}\) [day] & \# & \(t_{\textsc{bjd}} - t_{0}\) [day] & \# & \(t_{\textsc{bjd}} - t_{0}\) [day] & \# & \(t_{\textsc{bjd}} - t_{0}\) [day] \\[1pt]
    \specialrule{.4pt}{2pt}{0pt}
    1	&	\(330.1897_{-0.0034}^{+0.0034}\)	&	12	&	\(487.0041_{-0.0060}^{+0.0055}\)	&	25	&	\(672.3346_{-0.0031}^{+0.0030}\)	&	68	&	\(1285.2542_{-0.0015}^{+0.0015}\)	&	120	&	\(2026.5076_{-0.0015}^{+0.0015}\) \\
    2	&	\(344.4499_{-0.0048}^{+0.0049}\)	&	13	&	\(501.2493_{-0.0050}^{+0.0050}\)	&	43	&	\(928.9109_{-0.0026}^{+0.0028}\)	&	69	&	\(1299.5151_{-0.0019}^{+0.0019}\)	&	122	&	\(2055.0242_{-0.0019}^{+0.0019}\) \\
    5	&	\(387.2389_{-0.0051}^{+0.0051}\)	&	14	&	\(515.5045_{-0.0046}^{+0.0044}\)	&	47	&	\(985.9024_{-0.0045}^{+0.0042}\)	&	70	&	\(1313.7772_{-0.0015}^{+0.0014}\)	&	127	&	\(2126.2854_{-0.0021}^{+0.0020}\) \\
    6	&	\(401.5016_{-0.0033}^{+0.0032}\)	&	17	&	\(558.2707_{-0.0020}^{+0.0020}\)	&	49	&	\(1014.4174_{-0.0047}^{+0.0050}\)	&	71	&	\(1328.0406_{-0.0016}^{+0.0016}\)	&	128	&	\(2140.5308_{-0.0021}^{+0.0020}\) \\
    7	&	\(415.7570_{-0.0036}^{+0.0037}\)	&	18	&	\(572.5198_{-0.0020}^{+0.0020}\)	&	51	&	\(1042.9254_{-0.0018}^{+0.0017}\)	&	72	&	\(1342.3026_{-0.0014}^{+0.0014}\)	&	130	&	\(2169.0313_{-0.0014}^{+0.0014}\) \\
    8	&	\(430.0029_{-0.0035}^{+0.0035}\)	&	19	&	\(586.7907_{-0.0019}^{+0.0019}\)	&	52	&	\(1057.1874_{-0.0024}^{+0.0024}\)	&	73	&	\(1356.5535_{-0.0014}^{+0.0015}\)	&	131	&	\(2183.2800_{-0.0016}^{+0.0016}\) \\
    9	&	\(444.2563_{-0.0045}^{+0.0039}\)	&	20	&	\(601.0491_{-0.0019}^{+0.0019}\)	&	53	&	\(1071.4564_{-0.0028}^{+0.0028}\)	&	117	&	\(1983.7231_{-0.0012}^{+0.0013}\)	&	132	&	\(2197.5447_{-0.0023}^{+0.0023}\) \\
    10	&	\(458.5109_{-0.0037}^{+0.0037}\)	&	21	&	\(615.3137_{-0.0018}^{+0.0018}\)	&	65	&	\(1242.4828_{-0.0018}^{+0.0019}\)	&	118	&	\(1997.9843_{-0.0015}^{+0.0015}\)	&	146	&	\(2397.1076_{-0.0018}^{+0.0018}\) \\
    11	&	\(472.7612_{-0.0055}^{+0.0055}\)	&	24	&	\(658.0792_{-0.0038}^{+0.0038}\)	&	66	&	\(1256.7343_{-0.0020}^{+0.0018}\)	&	119	&	\(2012.2452_{-0.0016}^{+0.0015}\)	&	154	&	\(2511.1672_{-0.0016}^{+0.0017}\) \\
    \bottomrule
\end{tabularx}
\end{table}

\begin{table}[!hb]
    \centering
    \caption{Transit and limb-darkening parameters with median values and \(1\sigma\) uncertainties derived during \ttv\ extraction. The individual transit times for \toinumb\ are listed  in \autoref{table:TTVdataUpdated}.\label{tab:tableA1}}
    \begin{tabularx}{\textwidth}{l @{\hskip 5pt} r @{\hskip 17pt} l @{\hskip 5pt} r}
        \toprule
        \textsc{parameter [units]} & \textsc{value and} \(1\sigma\) \textsc{uncertainty} & \textsc{parameter} & \textsc{value and} \(1\sigma\) \textsc{uncertainty} \\[1pt]
        \specialrule{.4pt}{2pt}{0pt}
        \textsc{transit offset}, \textsc{tess ffi} [ppm] & \(119.07_{-81.18}^{+82.64}\) & \texttt{ld-quad-1}, \textsc{tess ffi} & \(0.2377_{-0.16}^{+0.17}\) \\ [1pt]
        \textsc{transit offset}, \textsc{tess} 2--min [ppm] & \(5.12_{-24.25}^{+24.80}\) & \texttt{ld-quad-2}, \textsc{tess ffi} & \(0.3498_{-0.21}^{+0.22}\) \\ [1pt]
        \textsc{transit offset}, Hazelwood [ppm] & \(909.09_{-347.67}^{+355.99}\) & \texttt{ld-quad-1}, \textsc{tess} 2--min & \(0.1501_{-0.10}^{+0.13}\) \\ [1pt]
        \textsc{transit offset}, \textsc{lco--sso} [ppm] & \(-65.90_{-197.73}^{+193.58}\) & \texttt{ld-quad-2}, \textsc{tess} 2--min & \(0.3051_{-0.17}^{+0.15}\) \\ [1pt]
        \textsc{transit offset}, Brierfield [ppm] & \(-51.86_{-448.81}^{+470.12}\) & \texttt{ld-quad-1}, \text{Hazelwood} & \(0.3253_{-0.22}^{+0.27}\) \\ [1pt]
        \textsc{transit offset}, \textsc{astep--r} [ppm] & \(363.68_{-169.04}^{+162.84}\) & \texttt{ld-quad-2}, \text{Hazelwood} & \(0.4535_{-0.28}^{+0.31}\) \\ [1pt]
        \textsc{transit offset}, \textsc{astep--b} [ppm] & \(-508.16_{-153.35}^{+150.01}\) & \texttt{ld-quad-1}, \textsc{lco--sso} & \(0.2728_{-0.18}^{+0.22}\) \\ [1pt]
        \textsc{transit jitter}, \textsc{tess ffi} [ppm] & \(1422.53_{-62.65}^{+66.54}\) & \texttt{ld-quad-2}, \textsc{lco--sso} & \(0.2676_{-0.19}^{+0.26}\) \\ [1pt]
        \textsc{transit jitter}, \textsc{tess} 2--min [ppm] & \(0.71_{-67.86}^{+67.14}\) & \texttt{ld-quad-1}, \text{Brierfield} & \(0.5890_{-0.30}^{+0.25}\) \\ [1pt]
        \textsc{transit jitter}, \text{Hazelwood} [ppm] & \(989.74_{-2525.61}^{+684.520}\) & \texttt{ld-quad-2}, \text{Brierfield} & \(0.5632_{-0.32}^{+0.28}\) \\ [1pt]
        \textsc{transit jitter}, \textsc{lco--sso} [ppm] & \(-7.42_{-214.24}^{+218.96}\) & \texttt{ld-quad-1}, \textsc{astep--r} & \(0.0842_{-0.06}^{+0.11}\) \\ [1pt]
        \textsc{transit jitter}, Brierfield [ppm] & \(61.18_{-558.60}^{+562.79}\) & \texttt{ld-quad-2}, \textsc{astep--r} & \(0.1894_{-0.13}^{+0.19}\) \\ [1pt]
        \textsc{transit jitter}, \textsc{astep--r} [ppm] & \(2567.67_{-119.45}^{+126.90}\) & \texttt{ld-quad-1}, \textsc{astep--b} & \(0.7601_{-0.13}^{+0.10}\) \\ [1pt]
        \textsc{transit jitter}, \textsc{astep--b} [ppm] & \(1353.28_{-133.08}^{+118.89}\) & \texttt{ld-quad-2}, \textsc{astep--b} & \(0.0819_{-0.06}^{+0.12}\) \\
        \bottomrule
    \end{tabularx}
\end{table}
\vspace*{-8pt}

\begin{table*}[!ht]
    \caption{Radial velocity and activity indices of \toinum\ measured with \feros. The times are all provided relative to an initial time \(t_{0} = 2458000.0\) days.\label{tab:FEROS_RVS}}
\begin{tabularx}{\textwidth}{r @{\hskip 26pt} r r r r r r r r r r r}
    \toprule
    \(t_{\bjd} - t_{0}\) [day] & \rv\ [km/s] & \(\sigma_{\rm{\rv}}\) [km/s] & \bis & \(\sigma_{\rm{\bis}}\) & \fwhm & \(H\!\!-\!\alpha\) & \(\sigma_{H-\alpha}\) & CA\,II & \(\sigma_{\rm{CA\,II}}\) & S\,MW & \(\sigma_{\rm{S\,MW}}\) \\ [1pt]
    \specialrule{.4pt}{2pt}{0pt}
    \(910.7216\) & \(9.6528\) & \(0.0091\) &  \(0.006\) & \(0.012\) & \(10.2012\) & \(-0.1923\) &      --- & \(0.1283\) & \(0.0153\) & \(0.1575\) & \(0.0176\) \\
    \(911.6443\) & \(9.6184\) & \(0.0087\) & \(-0.005\) & \(0.012\) & \(10.1725\) &  \(0.1174\) & \(0.0028\) & \(0.1664\) & \(0.0125\) & \(0.1999\) & \(0.0148\) \\
    \(917.7519\) & \(9.3161\) & \(0.0106\) & \(-0.202\) & \(0.013\) & \(10.4242\) &  \(0.1093\) & \(0.0032\) & \(0.2100\) & \(0.0260\) & \(0.2484\) & \(0.0295\) \\
    \(919.5835\) & \(9.4108\) & \(0.0154\) & \(-0.177\) & \(0.018\) & \(10.3973\) &  \(0.1740\) & \(0.0071\) & \(0.0921\) & \(0.0243\) & \(0.1172\) & \(0.0274\) \\
    \(925.6211\) & \(9.6405\) & \(0.0098\) &  \(0.047\) & \(0.013\) & \(10.1840\) &  \(0.1203\) & \(0.0034\) & \(0.1664\) & \(0.0141\) & \(0.1999\) & \(0.0165\) \\
    \(1188.7572\) & \(9.5400\) & \(0.0096\) & \(-0.018\) & \(0.013\) & \(10.1068\) &  \(0.1118\) &  \(0.003\) & \(0.1601\) & \(0.0147\) & \(0.1929\) & \(0.0171\) \\
    \(1189.7607\) & \(9.5333\) & \(0.0084\) & \(-0.007\) & \(0.012\) & \(10.0181\) &  \(0.1171\) & \(0.0026\) & \(0.1292\) & \(0.0106\) & \(0.1586\) & \(0.0128\) \\
    \(1190.7158\) & \(9.5301\) & \(0.0082\) & \(-0.024\) & \(0.011\) & \(10.0124\) &  \(0.1149\) & \(0.0024\) & \(0.1131\) & \(0.0134\) & \(0.1406\) & \(0.0156\) \\
    \(1192.7317\) & \(9.6021\) & \(0.0128\) & \(-0.025\) & \(0.016\) & \(10.2031\) &  \(0.1255\) & \(0.0053\) & \(0.1078\) & \(0.0201\) & \(0.1347\) & \(0.0228\) \\
    \(1194.7121\) & \(9.6394\) & \(0.0088\) &  \(0.022\) & \(0.012\) &  \(9.9833\) &  \(0.1165\) & \(0.0027\) & \(0.1352\) & \(0.0167\) & \(0.1652\) & \(0.0192\) \\
    \(1196.6930\) & \(9.6209\) & \(0.0080\) &  \(0.006\) & \(0.011\) &  \(9.9686\) &  \(0.1122\) & \(0.0026\) & \(0.1075\) & \(0.0095\) & \(0.1344\) & \(0.0115\) \\
    \(1206.7241\) & \(9.5703\) & \(0.0082\) & \(-0.025\) & \(0.011\) & \(10.0402\) &  \(0.1126\) & \(0.0026\) & \(0.1403\) & \(0.0096\) & \(0.1709\) & \(0.0118\) \\
    \(1209.7175\) & \(9.6037\) & \(0.0106\) & \(-0.011\) & \(0.014\) & \(10.0491\) &  \(0.1256\) & \(0.0034\) & \(0.1187\) & \(0.0194\) & \(0.1469\) & \(0.0221\) \\
    \(2030.5804\) & \(9.4737\) & \(0.0112\) & \(-0.056\) & \(0.015\) & \(10.2137\) &  \(0.1547\) & \(0.0042\) & \(0.1128\) & \(0.0186\) & \(0.1403\) & \(0.0212\) \\
    \(2031.6462\) & \(9.4784\) & \(0.0102\) & \(-0.013\) & \(0.013\) & \(10.1193\) &  \(0.1373\) & \(0.0036\) & \(0.1401\) & \(0.0187\) & \(0.1707\) & \(0.0213\) \\
    \(2033.6194\) & \(9.5323\) & \(0.0103\) &  \(0.056\) & \(0.013\) & \(10.1850\) &  \(0.1240\) & \(0.0032\) & \(0.2533\) & \(0.0209\) & \(0.2965\) & \(0.0241\) \\
    \(2035.6283\) & \(9.5684\) & \(0.0082\) & \(-0.026\) & \(0.011\) & \(10.1433\) & \(-0.0196\) &  \(0.001\) & \(0.0892\) & \(0.0115\) & \(0.1141\) & \(0.0136\) \\
    \(2064.6046\) & \(9.5504\) & \(0.0117\) &  \(0.001\) & \(0.015\) & \(10.2329\) &  \(0.1546\) & \(0.0045\) & \(0.1657\) & \(0.0292\) & \(0.1991\) & \(0.0329\) \\
    \(2066.5663\) & \(9.5775\) & \(0.0088\) & \(-0.030\) & \(0.012\) & \(10.0608\) &  \(0.1256\) & \(0.0028\) & \(0.0817\) & \(0.0163\) & \(0.1057\) & \(0.0187\) \\
    \(2100.5691\) & \(9.4427\) & \(0.0127\) &  \(0.029\) & \(0.015\) &  \(9.9596\) &  \(0.1230\) & \(0.0038\) & \(0.1650\) & \(0.0807\) & \(0.1983\) & \(0.0899\) \\
    \(2102.5615\) & \(9.4990\) & \(0.0104\) &  \(0.026\) & \(0.013\) & \(10.4111\) &  \(0.1341\) & \(0.0032\) & \(1.3270\) & \(0.1412\) & \(1.4901\) & \(0.1590\) \\
    \bottomrule
    \end{tabularx}
\end{table*}
\vspace*{-8pt}

\begin{table*}[ht!]
    \centering
    \caption{Transit times with uncertainties. The times are all provided relative to an initial time \(t_{0} = 2458000.0\) days.}
    \label{tab:transit_times}
    
    \begin{tabularx}{\textwidth}{r c @{\hskip 48pt} r c @{\hskip 48pt} r c @{\hskip 48pt} r c @{\hskip 48pt}}
    \toprule
    \# & \(t_{n} - t_{0}\) [BJD] & \# & \(t_{n} - t_{0}\) [BJD] & \# & \(t_{n} - t_{0}\) [BJD] & \# & \(t_{n} - t_{0}\) [BJD] \\
    \midrule
    189	&	\(3010.1676_{-0.1370}^{+0.1504}\)	&	239	&	\(3722.9157_{-0.1721}^{+0.1887}\)	&	289	&	\(4435.6652_{-0.2095}^{+0.2295}\)	&	339	&	\(5148.4124_{-0.2448}^{+0.2679}\) \\
    190	&	\(3024.4230_{-0.1394}^{+0.1536}\)	&	240	&	\(3737.1722_{-0.1770}^{+0.1945}\)	&	290	&	\(4449.9202_{-0.2127}^{+0.2340}\)	&	340	&	\(5162.6686_{-0.2512}^{+0.2751}\) \\
    191	&	\(3038.6768_{-0.1429}^{+0.1575}\)	&	241	&	\(3751.4258_{-0.1796}^{+0.1975}\)	&	291	&	\(4464.1743_{-0.2181}^{+0.2397}\)	&	341	&	\(5176.9221_{-0.2549}^{+0.2802}\) \\
    192	&	\(3052.9287_{-0.1449}^{+0.1603}\)	&	242	&	\(3765.6776_{-0.1835}^{+0.2025}\)	&	292	&	\(4478.4263_{-0.2212}^{+0.2442}\)	&	342	&	\(5191.1745_{-0.2589}^{+0.2864}\) \\
    193	&	\(3067.1789_{-0.1474}^{+0.1628}\)	&	243	&	\(3779.9283_{-0.1859}^{+0.2054}\)	&	293	&	\(4492.6772_{-0.2239}^{+0.2470}\)	&	343	&	\(5205.4256_{-0.2632}^{+0.2892}\) \\
    194	&	\(3081.4285_{-0.1493}^{+0.1647}\)	&	244	&	\(3794.1783_{-0.1876}^{+0.2067}\)	&	294	&	\(4506.9273_{-0.2269}^{+0.2495}\)	&	344	&	\(5219.6764_{-0.2650}^{+0.2910}\) \\
    195	&	\(3095.6783_{-0.1503}^{+0.1652}\)	&	245	&	\(3808.4278_{-0.1893}^{+0.2079}\)	&	295	&	\(4521.1778_{-0.2275}^{+0.2498}\)	&	345	&	\(5233.9269_{-0.2659}^{+0.2916}\) \\
    196	&	\(3109.9280_{-0.1509}^{+0.1650}\)	&	246	&	\(3822.6786_{-0.1892}^{+0.2069}\)	&	296	&	\(4535.4283_{-0.2270}^{+0.2480}\)	&	346	&	\(5248.1783_{-0.2649}^{+0.2896}\) \\
    197	&	\(3124.1795_{-0.1504}^{+0.1643}\)	&	247	&	\(3836.9298_{-0.1877}^{+0.2049}\)	&	297	&	\(4549.6804_{-0.2259}^{+0.2464}\)	&	347	&	\(5262.4307_{-0.2611}^{+0.2864}\) \\
    198	&	\(3138.4322_{-0.1480}^{+0.1624}\)	&	248	&	\(3851.1830_{-0.1857}^{+0.2037}\)	&	298	&	\(4563.9341_{-0.2202}^{+0.2431}\)	&	348	&	\(5276.6843_{-0.2578}^{+0.2848}\) \\
    199	&	\(3152.6865_{-0.1459}^{+0.1611}\)	&	249	&	\(3865.4386_{-0.1806}^{+0.2000}\)	&	299	&	\(4578.1888_{-0.2178}^{+0.2412}\)	&	349	&	\(5290.9405_{-0.2526}^{+0.2796}\) \\
    200	&	\(3166.9445_{-0.1425}^{+0.1571}\)	&	250	&	\(3879.6946_{-0.1791}^{+0.1982}\)	&	300	&	\(4592.4471_{-0.2139}^{+0.2363}\)	&	350	&	\(5305.1965_{-0.2507}^{+0.2779}\) \\
    201	&	\(3181.2018_{-0.1421}^{+0.1558}\)	&	251	&	\(3893.9551_{-0.1765}^{+0.1936}\)	&	301	&	\(4606.7041_{-0.2129}^{+0.2337}\)	&	351	&	\(5319.4563_{-0.2472}^{+0.2719}\) \\
    202	&	\(3195.4645_{-0.1397}^{+0.1543}\)	&	252	&	\(3908.2131_{-0.1753}^{+0.1933}\)	&	302	&	\(4620.9657_{-0.2106}^{+0.2321}\)	&	352	&	\(5333.7138_{-0.2471}^{+0.2711}\) \\
    203	&	\(3209.7234_{-0.1406}^{+0.1545}\)	&	253	&	\(3922.4759_{-0.1760}^{+0.1935}\)	&	303	&	\(4635.2237_{-0.2116}^{+0.2331}\)	&	353	&	\(5347.9749_{-0.2486}^{+0.2728}\) \\
    204	&	\(3223.9864_{-0.1431}^{+0.1566}\)	&	254	&	\(3936.7342_{-0.1774}^{+0.1950}\)	&	304	&	\(4649.4851_{-0.2152}^{+0.2354}\)	&	354	&	\(5362.2323_{-0.2505}^{+0.2741}\) \\
    205	&	\(3238.2443_{-0.1457}^{+0.1597}\)	&	255	&	\(3950.9947_{-0.1829}^{+0.2002}\)	&	305	&	\(4663.7421_{-0.2181}^{+0.2382}\)	&	355	&	\(5376.4915_{-0.2553}^{+0.2791}\) \\
    206	&	\(3252.5028_{-0.1500}^{+0.1651}\)	&	256	&	\(3965.2509_{-0.1858}^{+0.2035}\)	&	306	&	\(4677.9999_{-0.2229}^{+0.2444}\)	&	356	&	\(5390.7471_{-0.2590}^{+0.2830}\) \\
    207	&	\(3266.7575_{-0.1528}^{+0.1687}\)	&	257	&	\(3979.5065_{-0.1900}^{+0.2092}\)	&	307	&	\(4692.2544_{-0.2270}^{+0.2495}\)	&	357	&	\(5405.0026_{-0.2651}^{+0.2906}\) \\
    208	&	\(3281.0106_{-0.1560}^{+0.1720}\)	&	258	&	\(3993.7595_{-0.1937}^{+0.2132}\)	&	308	&	\(4706.5079_{-0.2318}^{+0.2551}\)	&	358	&	\(5419.2556_{-0.2686}^{+0.2957}\) \\
    209	&	\(3295.2620_{-0.1588}^{+0.1756}\)	&	259	&	\(4008.0109_{-0.1970}^{+0.2178}\)	&	309	&	\(4720.7595_{-0.2349}^{+0.2599}\)	&	359	&	\(5433.5077_{-0.2732}^{+0.3010}\) \\
    210	&	\(3309.5121_{-0.1609}^{+0.1775}\)	&	260	&	\(4022.2613_{-0.1995}^{+0.2201}\)	&	310	&	\(4735.0103_{-0.2377}^{+0.2616}\)	&	360	&	\(5447.7587_{-0.2772}^{+0.3038}\) \\
    211	&	\(3323.7616_{-0.1627}^{+0.1794}\)	&	261	&	\(4036.5114_{-0.2010}^{+0.2213}\)	&	311	&	\(4749.2604_{-0.2403}^{+0.2641}\)	&	361	&	\(5462.0096_{-0.2783}^{+0.3055}\) \\
    212	&	\(3338.0116_{-0.1634}^{+0.1794}\)	&	262	&	\(4050.7611_{-0.2023}^{+0.2215}\)	&	312	&	\(4763.5111_{-0.2403}^{+0.2632}\)	&	362	&	\(5476.2604_{-0.2784}^{+0.3048}\) \\
    213	&	\(3352.2617_{-0.1637}^{+0.1785}\)	&	263	&	\(4065.0123_{-0.2017}^{+0.2202}\)	&	313	&	\(4777.7622_{-0.2394}^{+0.2609}\)	&	363	&	\(5490.5122_{-0.2773}^{+0.3025}\) \\
    214	&	\(3366.5137_{-0.1626}^{+0.1776}\)	&	264	&	\(4079.2643_{-0.1992}^{+0.2178}\)	&	314	&	\(4792.0148_{-0.2370}^{+0.2593}\)	&	364	&	\(5504.7653_{-0.2713}^{+0.2991}\) \\
    215	&	\(3380.7673_{-0.1589}^{+0.1751}\)	&	265	&	\(4093.5180_{-0.1964}^{+0.2164}\)	&	315	&	\(4806.2694_{-0.2309}^{+0.2554}\)	&	365	&	\(5519.0194_{-0.2684}^{+0.2971}\) \\
    216	&	\(3395.0223_{-0.1569}^{+0.1737}\)	&	266	&	\(4107.7746_{-0.1917}^{+0.2122}\)	&	316	&	\(4820.5246_{-0.2287}^{+0.2533}\)	&	366	&	\(5533.2763_{-0.2637}^{+0.2918}\) \\
    217	&	\(3409.2812_{-0.1540}^{+0.1691}\)	&	267	&	\(4122.0310_{-0.1904}^{+0.2100}\)	&	317	&	\(4834.7835_{-0.2250}^{+0.2480}\)	&	367	&	\(5547.5325_{-0.2617}^{+0.2900}\) \\
    218	&	\(3423.5389_{-0.1531}^{+0.1681}\)	&	268	&	\(4136.2920_{-0.1874}^{+0.2060}\)	&	318	&	\(4849.0408_{-0.2241}^{+0.2460}\)	&	368	&	\(5561.7927_{-0.2585}^{+0.2843}\) \\
    219	&	\(3437.8018_{-0.1515}^{+0.1670}\)	&	269	&	\(4150.5502_{-0.1868}^{+0.2065}\)	&	319	&	\(4863.3023_{-0.2230}^{+0.2455}\)	&	369	&	\(5576.0501_{-0.2593}^{+0.2840}\) \\
    220	&	\(3452.0606_{-0.1528}^{+0.1678}\)	&	270	&	\(4164.8126_{-0.1886}^{+0.2073}\)	&	320	&	\(4877.5602_{-0.2249}^{+0.2465}\)	&	370	&	\(5590.3109_{-0.2613}^{+0.2866}\) \\
    221	&	\(3466.3229_{-0.1560}^{+0.1708}\)	&	271	&	\(4179.0705_{-0.1910}^{+0.2090}\)	&	321	&	\(4891.8208_{-0.2287}^{+0.2498}\)	&	371	&	\(5604.5679_{-0.2638}^{+0.2885}\) \\
    222	&	\(3480.5802_{-0.1590}^{+0.1743}\)	&	272	&	\(4193.3301_{-0.1963}^{+0.2150}\)	&	322	&	\(4906.0774_{-0.2312}^{+0.2530}\)	&	372	&	\(5618.8264_{-0.2693}^{+0.2943}\) \\
    223	&	\(3494.8376_{-0.1633}^{+0.1798}\)	&	273	&	\(4207.5857_{-0.1992}^{+0.2185}\)	&	323	&	\(4920.3344_{-0.2368}^{+0.2598}\)	&	373	&	\(5633.0815_{-0.2735}^{+0.2986}\) \\
    224	&	\(3509.0918_{-0.1664}^{+0.1834}\)	&	274	&	\(4221.8405_{-0.2039}^{+0.2240}\)	&	324	&	\(4934.5884_{-0.2413}^{+0.2645}\)	&	374	&	\(5647.3363_{-0.2786}^{+0.3063}\) \\
    225	&	\(3523.3442_{-0.1695}^{+0.1871}\)	&	275	&	\(4236.0930_{-0.2078}^{+0.2286}\)	&	325	&	\(4948.8413_{-0.2451}^{+0.2708}\)	&	375	&	\(5661.5889_{-0.2832}^{+0.3116}\) \\
    226	&	\(3537.5952_{-0.1727}^{+0.1908}\)	&	276	&	\(4250.3441_{-0.2103}^{+0.2323}\)	&	326	&	\(4963.0926_{-0.2491}^{+0.2745}\)	&	376	&	\(5675.8408_{-0.2873}^{+0.3155}\) \\
    227	&	\(3551.8452_{-0.1742}^{+0.1920}\)	&	277	&	\(4264.5943_{-0.2132}^{+0.2348}\)	&	327	&	\(4977.3434_{-0.2515}^{+0.2763}\)	&	377	&	\(5690.0917_{-0.2909}^{+0.3185}\) \\
    228	&	\(3566.0947_{-0.1761}^{+0.1938}\)	&	278	&	\(4278.8445_{-0.2144}^{+0.2357}\)	&	328	&	\(4991.5936_{-0.2535}^{+0.2783}\)	&	378	&	\(5704.3428_{-0.2912}^{+0.3198}\) \\
    229	&	\(3580.3450_{-0.1764}^{+0.1933}\)	&	279	&	\(4293.0946_{-0.2149}^{+0.2348}\)	&	329	&	\(5005.8447_{-0.2530}^{+0.2765}\)	&	379	&	\(5718.5940_{-0.2907}^{+0.3177}\) \\
    230	&	\(3594.5956_{-0.1757}^{+0.1918}\)	&	280	&	\(4307.3463_{-0.2139}^{+0.2334}\)	&	330	&	\(5020.0963_{-0.2508}^{+0.2737}\)	&	380	&	\(5732.8463_{-0.2885}^{+0.3154}\) \\
    231	&	\(3608.8482_{-0.1746}^{+0.1908}\)	&	281	&	\(4321.5991_{-0.2098}^{+0.2307}\)	&	331	&	\(5034.3495_{-0.2474}^{+0.2723}\)	&	381	&	\(5747.1001_{-0.2818}^{+0.3117}\) \\
    232	&	\(3623.1028_{-0.1697}^{+0.1877}\)	&	282	&	\(4335.8533_{-0.2070}^{+0.2289}\)	&	332	&	\(5048.6049_{-0.2416}^{+0.2676}\)	&	382	&	\(5761.3546_{-0.2790}^{+0.3092}\) \\
    233	&	\(3637.3583_{-0.1680}^{+0.1860}\)	&	283	&	\(4350.1108_{-0.2028}^{+0.2245}\)	&	333	&	\(5062.8605_{-0.2397}^{+0.2656}\)	&	383	&	\(5775.6122_{-0.2746}^{+0.3041}\) \\
    234	&	\(3651.6181_{-0.1654}^{+0.1813}\)	&	284	&	\(4364.3675_{-0.2017}^{+0.2218}\)	&	334	&	\(5077.1199_{-0.2361}^{+0.2598}\)	&	384	&	\(5789.8687_{-0.2726}^{+0.3018}\) \\
    235	&	\(3665.8760_{-0.1641}^{+0.1805}\)	&	285	&	\(4378.6289_{-0.1987}^{+0.2186}\)	&	335	&	\(5091.3773_{-0.2352}^{+0.2584}\)	&	385	&	\(5804.1289_{-0.2706}^{+0.2970}\) \\
    236	&	\(3680.1390_{-0.1634}^{+0.1801}\)	&	286	&	\(4392.8870_{-0.1989}^{+0.2198}\)	&	336	&	\(5105.6387_{-0.2361}^{+0.2594}\)	&	386	&	\(5818.3863_{-0.2717}^{+0.2978}\) \\
    237	&	\(3694.3975_{-0.1653}^{+0.1813}\)	&	287	&	\(4407.1490_{-0.2013}^{+0.2213}\)	&	337	&	\(5119.8964_{-0.2375}^{+0.2601}\)	&	387	&	\(5832.6465_{-0.2748}^{+0.3006}\) \\
    238	&	\(3708.6590_{-0.1693}^{+0.1854}\)	&	288	&	\(4421.4065_{-0.2048}^{+0.2234}\)	&	338	&	\(5134.1563_{-0.2416}^{+0.2644}\)	&	388	&	\(5846.9032_{-0.2775}^{+0.3032}\) \\
    \bottomrule
\end{tabularx}
\end{table*}

\addtocounter{table}{-1}
\begin{table*}[ht!]
\centering
\caption{Transit times with uncertainties.}
    \begin{tabularx}{\textwidth}{r c @{\hskip 48pt} r c @{\hskip 48pt} r c @{\hskip 48pt} r c @{\hskip 48pt}}
    \toprule
    \# & \(t_{n} - t_{0}\) [BJD] & \# & \(t_{n} - t_{0}\) [BJD] & \# & \(t_{n} - t_{0}\) [BJD] & \# & \(t_{n} - t_{0}\) [BJD] \\
    \midrule
    389	&	\(5861.1609_{-0.2835}^{+0.3095}\)	&	418	&	\(6274.5408_{-0.2944}^{+0.3255}\)	&	447	&	\(6687.9305_{-0.3331}^{+0.3679}\)	&	476	&	\(7101.3360_{-0.3644}^{+0.3988}\) \\
    390	&	\(5875.4156_{-0.2874}^{+0.3143}\)	&	419	&	\(6288.8009_{-0.2957}^{+0.3235}\)	&	448	&	\(6702.1845_{-0.3297}^{+0.3660}\)	&	477	&	\(7115.5878_{-0.3662}^{+0.4006}\) \\
    391	&	\(5889.6699_{-0.2924}^{+0.3223}\)	&	420	&	\(6303.0579_{-0.2976}^{+0.3262}\)	&	449	&	\(6716.4410_{-0.3249}^{+0.3599}\)	&	478	&	\(7129.8398_{-0.3667}^{+0.4027}\) \\
    392	&	\(5903.9221_{-0.2978}^{+0.3275}\)	&	421	&	\(6317.3171_{-0.3021}^{+0.3296}\)	&	450	&	\(6730.6967_{-0.3225}^{+0.3581}\)	&	479	&	\(7144.0922_{-0.3643}^{+0.4004}\) \\
    393	&	\(5918.1739_{-0.3014}^{+0.3302}\)	&	422	&	\(6331.5730_{-0.3056}^{+0.3333}\)	&	451	&	\(6744.9557_{-0.3183}^{+0.3520}\)	&	480	&	\(7158.3452_{-0.3605}^{+0.3975}\) \\
    394	&	\(5932.4248_{-0.3042}^{+0.3331}\)	&	423	&	\(6345.8294_{-0.3110}^{+0.3407}\)	&	452	&	\(6759.2125_{-0.3193}^{+0.3503}\)	&	481	&	\(7172.5998_{-0.3539}^{+0.3927}\) \\
    395	&	\(5946.6762_{-0.3040}^{+0.3332}\)	&	424	&	\(6360.0832_{-0.3156}^{+0.3459}\)	&	453	&	\(6773.4721_{-0.3225}^{+0.3521}\)	&	482	&	\(7186.8544_{-0.3515}^{+0.3900}\) \\
    396	&	\(5960.9278_{-0.3025}^{+0.3304}\)	&	425	&	\(6374.3366_{-0.3215}^{+0.3533}\)	&	454	&	\(6787.7286_{-0.3247}^{+0.3549}\)	&	483	&	\(7201.1118_{-0.3463}^{+0.3845}\) \\
    397	&	\(5975.1806_{-0.2988}^{+0.3282}\)	&	426	&	\(6388.5884_{-0.3266}^{+0.3577}\)	&	455	&	\(6801.9866_{-0.3298}^{+0.3597}\)	&	484	&	\(7215.3679_{-0.3439}^{+0.3829}\) \\
    398	&	\(5989.4351_{-0.2921}^{+0.3241}\)	&	427	&	\(6402.8401_{-0.3288}^{+0.3594}\)	&	456	&	\(6816.2417_{-0.3331}^{+0.3645}\)	&	485	&	\(7229.6270_{-0.3441}^{+0.3769}\) \\
    399	&	\(6003.6900_{-0.2899}^{+0.3213}\)	&	428	&	\(6417.0913_{-0.3295}^{+0.3618}\)	&	457	&	\(6830.4970_{-0.3394}^{+0.3722}\)	&	486	&	\(7243.8835_{-0.3456}^{+0.3767}\) \\
    400	&	\(6017.9481_{-0.2853}^{+0.3164}\)	&	429	&	\(6431.3433_{-0.3284}^{+0.3592}\)	&	458	&	\(6844.7500_{-0.3456}^{+0.3779}\)	&	487	&	\(7258.1424_{-0.3493}^{+0.3812}\) \\
    401	&	\(6032.2047_{-0.2833}^{+0.3136}\)	&	430	&	\(6445.5961_{-0.3228}^{+0.3555}\)	&	459	&	\(6859.0029_{-0.3504}^{+0.3840}\)	&	488	&	\(7272.3983_{-0.3520}^{+0.3842}\) \\
    402	&	\(6046.4650_{-0.2833}^{+0.3100}\)	&	431	&	\(6459.8497_{-0.3195}^{+0.3535}\)	&	460	&	\(6873.2546_{-0.3542}^{+0.3865}\)	&	489	&	\(7286.6552_{-0.3568}^{+0.3905}\) \\
    403	&	\(6060.7222_{-0.2842}^{+0.3119}\)	&	432	&	\(6474.1056_{-0.3140}^{+0.3479}\)	&	461	&	\(6887.5065_{-0.3540}^{+0.3884}\)	&	490	&	\(7300.9095_{-0.3623}^{+0.3949}\) \\
    404	&	\(6074.9820_{-0.2884}^{+0.3149}\)	&	433	&	\(6488.3611_{-0.3118}^{+0.3457}\)	&	462	&	\(6901.7584_{-0.3534}^{+0.3876}\)	&	491	&	\(7315.1639_{-0.3692}^{+0.4040}\) \\
    405	&	\(6089.2382_{-0.2914}^{+0.3182}\)	&	434	&	\(6502.6199_{-0.3067}^{+0.3401}\)	&	463	&	\(6916.0111_{-0.3504}^{+0.3849}\)	&	492	&	\(7329.4165_{-0.3748}^{+0.4101}\) \\
    406	&	\(6103.4953_{-0.2974}^{+0.3251}\)	&	435	&	\(6516.8767_{-0.3066}^{+0.3378}\)	&	464	&	\(6930.2651_{-0.3432}^{+0.3803}\)	&	493	&	\(7343.6692_{-0.3783}^{+0.4133}\) \\
    407	&	\(6117.7495_{-0.3012}^{+0.3300}\)	&	436	&	\(6531.1366_{-0.3089}^{+0.3377}\)	&	465	&	\(6944.5194_{-0.3405}^{+0.3781}\)	&	494	&	\(7357.9210_{-0.3796}^{+0.4152}\) \\
    408	&	\(6132.0033_{-0.3070}^{+0.3379}\)	&	437	&	\(6545.3934_{-0.3112}^{+0.3406}\)	&	466	&	\(6958.7764_{-0.3357}^{+0.3722}\)	&	495	&	\(7372.1733_{-0.3789}^{+0.4165}\) \\
    409	&	\(6146.2553_{-0.3123}^{+0.3429}\)	&	438	&	\(6559.6520_{-0.3161}^{+0.3445}\)	&	467	&	\(6973.0323_{-0.3331}^{+0.3705}\)	&	496	&	\(7386.4261_{-0.3743}^{+0.4129}\) \\
    410	&	\(6160.5070_{-0.3152}^{+0.3448}\)	&	439	&	\(6573.9074_{-0.3193}^{+0.3488}\)	&	468	&	\(6987.2914_{-0.3309}^{+0.3643}\)	&	497	&	\(7400.6795_{-0.3707}^{+0.4101}\) \\
    411	&	\(6174.7580_{-0.3165}^{+0.3476}\)	&	440	&	\(6588.1633_{-0.3248}^{+0.3563}\)	&	469	&	\(7001.5481_{-0.3322}^{+0.3633}\)	&	498	&	\(7414.9347_{-0.3648}^{+0.4051}\) \\
    412	&	\(6189.0096_{-0.3161}^{+0.3462}\)	&	441	&	\(6602.4167_{-0.3306}^{+0.3619}\)	&	470	&	\(7015.8073_{-0.3356}^{+0.3667}\)	&	499	&	\(7429.1895_{-0.3623}^{+0.4022}\) \\
    413	&	\(6203.2619_{-0.3127}^{+0.3429}\)	&	442	&	\(6616.6698_{-0.3360}^{+0.3687}\)	&	471	&	\(7030.0635_{-0.3385}^{+0.3694}\)	&	500	&	\(7443.4473_{-0.3570}^{+0.3970}\) \\
    414	&	\(6217.5150_{-0.3091}^{+0.3409}\)	&	443	&	\(6630.9215_{-0.3405}^{+0.3722}\)	&	472	&	\(7044.3210_{-0.3431}^{+0.3751}\)	&	501	&	\(7457.7034_{-0.3553}^{+0.3948}\) \\
    415	&	\(6231.7703_{-0.3030}^{+0.3360}\)	&	444	&	\(6645.1733_{-0.3414}^{+0.3740}\)	&	473	&	\(7058.5757_{-0.3472}^{+0.3798}\)	&	502	&	\(7471.9624_{-0.3572}^{+0.3897}\) \\
    416	&	\(6246.0255_{-0.3009}^{+0.3334}\)	&	445	&	\(6659.4248_{-0.3415}^{+0.3748}\)	&	474	&	\(7072.8305_{-0.3543}^{+0.3881}\)	&	503	&	\(7486.2188_{-0.3592}^{+0.3903}\) \\
    417	&	\(6260.2840_{-0.2960}^{+0.3284}\)	&	446	&	\(6673.6771_{-0.3400}^{+0.3721}\)	&	475	&	\(7087.0833_{-0.3604}^{+0.3940}\)	&	504	&	\(7500.4772_{-0.3630}^{+0.3958}\) \\
    \bottomrule
\end{tabularx}
\end{table*}

\end{document}